\documentclass[pra,aps,10pt,superscriptaddress,twocolumn,floatfix]{revtex4-1}  
\usepackage{graphicx,color}
\usepackage{amsmath,amssymb,bm}
\usepackage{braket}		

\DeclareMathOperator{\Tr}{Tr}

\usepackage[plainpages=false,pdfpagelabels,colorlinks=true,linkcolor=red,urlcolor=blue,citecolor=blue,pdftitle={Titl}e,pdfauthor={},pdfdisplaydoctitle=true,pdfduplex=DuplexFlipLongEdge]{hyperref}

\definecolor{darkred}{rgb}{0.90,0.2,0.2}
\definecolor{darkgreen}{rgb}{0,0.60,.2}
\definecolor{darkblue}{rgb}{0.1,0.3,1}
\definecolor{grey}{cmyk}{0,0,0,0.25}
\definecolor{orange}{cmyk}{0,0.6,0.8,0}

\begin{document}
\title{Quantum dynamics of impenetrable SU($N$) fermions in one-dimensional lattices}

\author{Yicheng Zhang}
\affiliation{Department of Physics, The Pennsylvania State University, University Park, Pennsylvania 16802, USA}
\author{Lev Vidmar}
\affiliation{Department of Theoretical Physics, J. Stefan Institute, SI-1000 Ljubljana, Slovenia}
\author{Marcos Rigol}
\affiliation{Department of Physics, The Pennsylvania State University, University Park, Pennsylvania 16802, USA}

\begin{abstract}
We study quantum quench dynamics in the Fermi-Hubbard model, and its SU($N$) generalizations, in one-dimensional lattices in the limit of infinite onsite repulsion between all flavors. We consider families of initial states with generalized Neel order, namely, initial state in which there is a periodic $N$-spin pattern with consecutive fermions carrying distinct spin flavors. We introduce an exact approach to describe the quantum evolution of those systems, and study two unique transient phenomena that occur during expansion dynamics in finite lattices. The first one is the dynamical emergence of Gaussian one-body correlations during the melting of sharp (generalized) Neel domain walls. Those correlations resemble the ones in the ground state of the SU($N$) model constrained to the same spin configurations. This is explained using an emergent eigenstate solution to the quantum dynamics. The second phenomenon is the transformation of the quasimomentum distribution of the expanding strongly interacting SU($N$) gas into the rapidity distribution after long times. Finally, we study equilibration in SU($N$) gasses and show that observables after equilibration are described by a generalized Gibbs ensemble. Our approach can be used to benchmark analytical and numerical calculations of dynamics of strongly correlated SU($N$) fermions at large $U$.
\end{abstract}
\maketitle


\section{Introduction} \label{sec1}

Nonequilibrium dynamics of isolated many-body quantum systems is currently one of the most active research fields in many-body quantum physics~\cite{polkovnikov_sengupta_review_11, eisert_friesdorf_review_15, dalessio_kafri_16}. A special class of systems, called integrable systems, represent an important cornerstone in this field~\cite{Calabrese_2016, essler_fagotti_2016, calabrese_cardy_2016, cazalilla_chung_2016, bernard_doyon_2016, caux_2016, vidmar16, ilievski_medenjak_2016, langen_gasenzer_2016, vasseur_moore_2016, deluca_mussardo_2016}. They possess an extensive number of (quasi-local) conserved quantities, and after equilibration expectation values of few-body observables can be described by generalized Gibbs ensembles (GGEs)~\cite{rigol_dunjko_07,vidmar16}. Several experimental groups have been exploring many-body quantum dynamics with ultracold atoms close to integrability~\cite{kinoshita_wenger_06, gring_kuhnert_12, ronzheimer13, meinert13, fukuhara13a, fukuhara13b, hild_fukuhara_14, vidmar15, langen_erne_15,scherg_18, tang_kao_18}.

In this work we study the paradigmatic one-dimensional (1D) Fermi-Hubbard model~\cite{essler-book} and its generalized 1D SU($N$) versions~\cite{cazalilla_2009, Gorshkov_2010, Taie_2012, Pagano_2014, Scazza_2014, Zhang_2014, cazalilla_rey_14, nataf_mila_14, volosniev_14, dufour_nataf_15, decamp_armagnat_16, decamp_junemann_16, laird_shi_17, jen_yip_18,zinner_18}. The Fermi-Hubbard model is a minimal model used to describe correlated electrons in solids, and provides an accurate description of ongoing experiments with ultracold fermions in optical lattices~\cite{esslinger_review_10}. After early theoretical studies of quantum quenches in the Fermi-Hubbard model in dimensions higher than one~\cite{moeckel_kehrein_08, moeckel_kehrein_09, eckstein_kollar_09, eckstein_kollar_10, schiro_fabrizio_10, schiro_fabrizio_11}, recent works have focused on quantum quenches in the 1D Fermi-Hubbard model~\cite{hm08, hm09, kajala11, langer12, bolech12, hamerla_uhrig_13, hamerla_uhrig_13_b, iyer_mondaini_14, riegger_orso_15, mei16, yin_radzihovsky_16, schlunzen_joost_17, bertini_tartaglia_2017, bleicker_uhrig_18}, which is a quintessential integrable model~\cite{essler-book}.

We introduce an exact approach to study quantum dynamics of 1D SU($N$) fermions, of which the Fermi-Hubbard model is the $N=2$ case, in the limit of infinite onsite repulsion. A key requirement of our approach is that the states of which we study the quantum evolution must have consecutive impenetrable SU($N$) fermions with different spin flavors. In those configurations, since the spin order is preserved at all times, one can think of each fermion as a {\it distinguishable quantum particle}. This allows us to use spin-charge separation to treat the charge degrees of freedom as spinless fermions, and the spin degrees of freedom by means of nonlocal strings of operators acting on spinless fermion wave functions. Using this decomposition, we develop an efficient (polynomial-time) way to compute one-body correlations. The ideas behind this approach were recently used to describe equilibrium properties of impenetrable 1D SU($N$) fermions~\cite{zhang18}, and can be viewed as a fermionic generalization of the Jordan-Wigner transformation, which maps spin-$\frac{1}{2}$ models onto spinless fermions~\cite{jordan_wigner_28, lieb61}. We focus on dynamics of states that, in addition of having consecutive fermions with distinct spin flavors, have periodic $N$-spin patterns that we call {\it generalized Neel order}. 

\begin{figure}[t]
\begin{center}
\includegraphics[width=0.99\columnwidth]{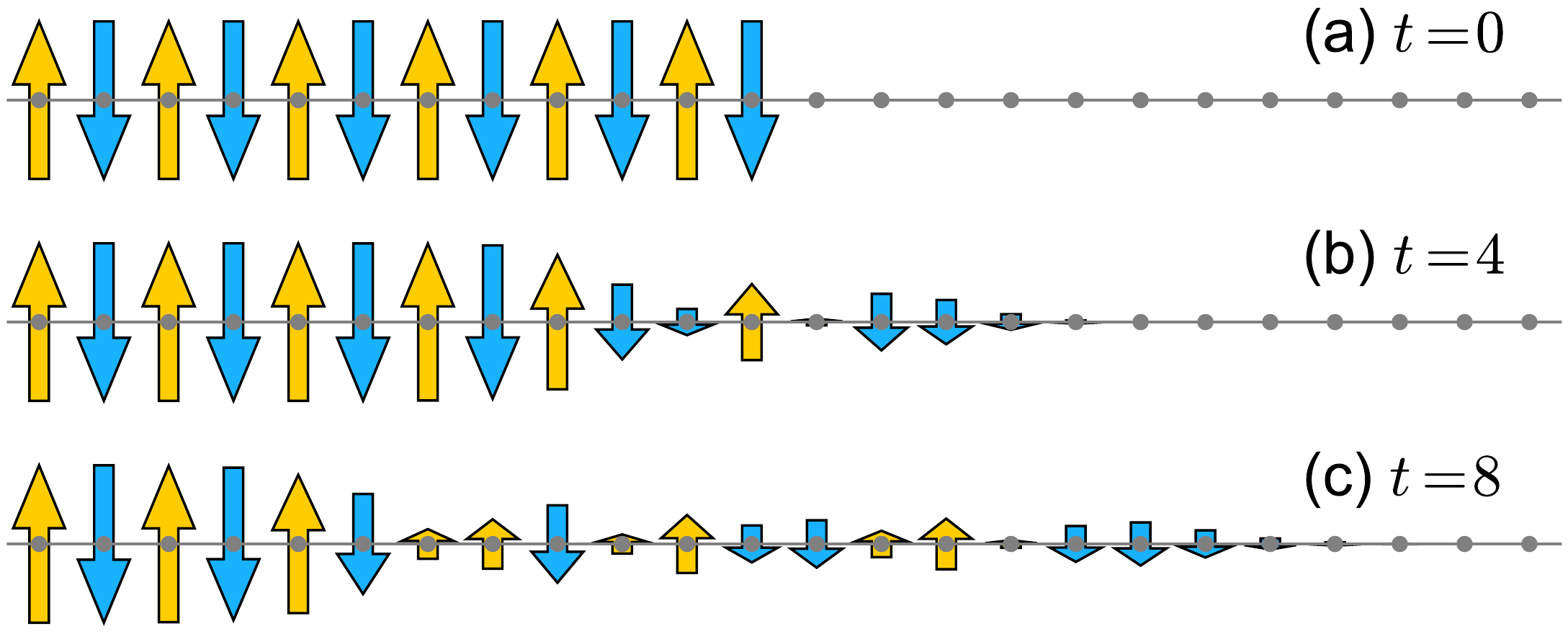}
\caption{Melting of a Neel domain wall in the Fermi-Hubbard model at infinite onsite repulsion. (a) At $t=0$ the system is in a Neel product state with 12 fermions. At $t>0$ the fermions expand into an empty lattice. The length of the arrows is proportional to the magnitude of the $z$-component of the spin at site $l$, $S^z_l(t) = |n^{\uparrow}(l;t) - n^{\downarrow}(l;t)|/2$, where $n^\sigma(l;t)$ are the site occupations of fermions with ``flavor'' $\sigma$ at time $t$. We show results at (b) $t=4$ and (c) $t=8$. This setup is studied in Sec.~\ref{sec3}.} \label{fig1}
\end{center}
\end{figure}

As a first application of our approach, we study the expansion (``melting'') of generalized Neel domain walls such as the one shown in Fig.~\ref{fig1} for the SU(2) case. This type of setup has been widely studied in the literature, in particular for lattice models without internal degrees of freedom~\cite{antal99, karevski02, ogata02, hunyadi04, rigol04, rigol05a, gobert05, antal08, eisler09, lancaster10, mossel10, santos11, eisler13, sabetta_misguich_13, alba14, hauschild15, vasseur15, lancaster16, bertini16, eisler_maislinger_16, hauschild_heidrichmeisner_16, vidmar_iyer_17, deluca_collura_17, ljubotina_znidaric_17, kormos_17, ribeiro17, herbrych_feiguin_17, stephan_17, eisler_bauernfeind_17, fagotti_17,lorenzo_17, bastianello_deluca_18, collura_deluca_18, bertini_piroli_18, bertini_18, denardis_bernard_18, alba_18, alba_19, mastyan_bertini_19, ljubotina_sotiriadis_19, mazza_perfetto_18}. We show that the resulting current-carrying state exhibits a dynamical emergence of one-body correlations with a Gaussian decay. These correlations resemble the ones in the ground state of the same model, constrained to the same spin configurations~\cite{zhang18}. We explain this observation using an emergent eigenstate solution to quantum dynamics~\cite{vidmar_iyer_17}, combined with the local density approximation.

As a second application, we study the expansion of harmonically trapped SU($N$) fermions after suddenly turning off the trap. Those quenches provide a fertile playground to unveil remarkable properties of correlated many-body systems~\cite{sutherland_98, rigol04, rigol05, rodriguez_manmana_06, kajala11, bolech12, vidmar13, boschi14, vidmar15, ren_wu_15, mei16, vidmar_iyer_17, xu_rigol_17, vidmar_xu_17, siegl_mistakidis_18, zhang_meurice_19}. They can be viewed as nontrivial time-of-flight expansions that occur in the presence of interactions. Here we demonstrate that, during the expansion, the quasimomentum distribution of impenetrable SU($N$) fermions evolves into the rapidity distribution, which is the initial quasimomentum distribution of the underlying spinless fermions to which the impenetrable SU($N$) fermions are mapped. Transformations of quasimomentum distributions of expanding fermions into rapidity distributions have been previously discussed in the context of the Fermi-Hubbard model~\cite{bolech12, mei16}. 

As a final application of our approach, we study the equilibration dynamics of SU($N$) fermions in a box trap. In that case, the quasimomentum distribution function after equilibration is described by means of a GGE in which the conserved quantities are the occupations of the single-particle eigenstates of the underlying spinless fermions to which the impenetrable SU($N$) fermions are mapped. This shows that the GGE can be used to describe observables after equilibration in integrable multiflavor fermionic systems with infinite onsite repulsion.

In addition to studying phenomena of relevance to current experiments with ultracold quantum gases, the approach introduced here can also be used to benchmark analytical and numerical calculations of quantum dynamics of strongly correlated SU($N$) fermions at large $U$, and specifically of the integrable SU(2) Fermi-Hubbard model. Quantum dynamics of integrable models non-mappable onto noninteracting ones have proved challenging. Analytic progress has been made in the context of the spin-$\frac{1}{2}$ XXZ chain, for which exact steady-state properties have been obtained for some classes of initial states either using special properties of those states (e.g, within the quench action approach~\cite{caux_essler_13, wouters_denardis_14, brockmann_wouters_14, pozsgay_mestyan_14, mestyan_pozsgay_15, alba_calabrese_16, caux_2016} and the quantum transfer matrix formalism~\cite{piroli_pozsgay_2017_a, piroli_pozsgay_2017_b}), or using the (quasi-local) conserved quantities of the model~\cite{fagotti_14, ilievski15, ilievski_quinn_2016, piroli_vernier_16, ilievski_quinn_17, piroli_vernier_17, pozsgay_vernier_17}. More recently, interest has grown in understanding systems with internal degrees of freedom, so-called nested systems, such as the Fermi-Hubbard model. In those systems, exact steady-state properties can also be obtained for a few classes of initial states. Examples include the two-component Lieb-Liniger gas~\cite{robinson_caux_16, robinson_caux_16_b}, the Fermi-Hubbard model~\cite{bolech12, mei16, bertini_tartaglia_2017,enej_17}, and the Lai-Sutherland model~\cite{mestyan_bertini_2017, piroli_vernier_19_a, piroli_vernier_19_b}. Nevertheless, their dynamics remains widely unexplored, and a systematic study of the GGE in such systems remains challenging due to difficulties in obtaining the complete set of conserved quantities.

The rest of the paper is organized as follows. In Sec.~\ref{sec2}, we introduce the nonequilibrium setup and the exact approach used to describe quantum dynamics after quenches. We then discuss two applications to transient phenomena: we study the melting of generalized Neel domain walls in Sec.~\ref{sec3}, and the sudden expansion of initially trapped systems in Sec.~\ref{sec4}. In Sec.~\ref{sec5}, we discuss equilibration to the GGE. A summary of our results is presented in Sec.~\ref{sec_conclusion}.

\section{Setup and quantum dynamics} \label{sec2}

Here, we discuss general considerations about the systems studied and introduce the exact approach developed to describe their dynamics.

\subsection{Distinguishable quantum particles}

We consider a generalized 1D Fermi-Hubbard model for SU($N$) fermions, $N$ is the number of flavors, with infinite onsite repulsion. For open chains with $L$ sites (on which we focus here), the Hamiltonian can be written as
\begin{equation}\label{def_H}
 \hat H_N = -J\sum_{l=1}^{L-1}\sum_{\sigma=1}^{N}\left[\hat f^{(\sigma)\dagger}_l \hat f^{(\sigma)}_{l+1}+\hat f^{(\sigma)\dagger}_{l+1} \hat f^{(\sigma)}_l\right]\,,
\end{equation}
where $\sigma$ is the spin flavor ($\sigma\in\{1,\dots,N\}$ in our notation). Infinite onsite repulsion is enforced by the constraints $\hat f^{(\sigma)\dagger}_l \hat f^{\dagger(\sigma')}_l = \hat f^{(\sigma)}_l \hat f^{(\sigma')}_l=0$, where $\hat f^{(\sigma)\dagger}_l$ ($\hat f^{(\sigma)}_l$) is the creation (annihilation) operator of a fermion with flavor $\sigma$ at site $l$. The traditional Fermi-Hubbard Hamiltonian corresponds to $N=2$. We set the hopping amplitude $J$ and the lattice spacing to unity.

An important property of Hamiltonian $\hat H_N$, which is a consequence of infinite onsite repulsion, is that it preserves the order of spin flavor configurations $\underline{\sigma}= \{ \sigma_1, \dots, \sigma_{N_p} \}$, where $\sigma_j \in \{ 1, \dots, N \}$, and $N_p$ is the number of particles. In a sector with $\underline{\sigma}$, any wave function in the occupation number basis is a linear combination of base kets $|\varphi_{\underline{x},\underline{\sigma}} \rangle = \prod_{j=1}^{N_p} \hat f_{x_j}^{(\sigma_j)\dagger} | \emptyset \rangle$, where every base ket denotes a different set of occupied sites $\underline{x} = \{ x_1, \dots, x_{N_p} \}$, with $x_j \in \{1, \dots, L \} $ and $x_1 < x_2 < \cdots < x_{N_p}$.

Here we are interested in quantum dynamics of initial states with a given spin configuration $\underline{\sigma}$. The most general states for which our approach is applicable are states with $\underline{\sigma}$ such that every pair of consecutive fermions carries distinct spin flavors
\begin{equation} \label{def_sigma_dqp}
 \underline{\sigma} = \{ \{\sigma_j \}; \; j=1,...,N_p; \; \sigma_j \neq \sigma_{j+1} \; \forall \, j < N_p \} \,.
\end{equation}
We call the impenetrable SU($N$) model governed by $\hat H_N$~(\ref{def_H}), constrained to a single spin configuration $\underline{\sigma}$ that obeys Eq.~(\ref{def_sigma_dqp}), a model of distinguishable quantum particles (DQPs).

In the SU(2) case, the DQP model is the traditional Fermi-Hubbard model within the sectors with states that have alternating spin flavors. An example of such an spin ordering is shown in Fig.~\ref{fig1}(a). In the SU($N$) case with $N>2$, different realizations of spin configurations satisfying Eq.~(\ref{def_sigma_dqp}) are possible. For simplicity, we focus on periodic $N$-spin patterns with consecutive fermions carrying distinct spin flavors, as described by
\begin{equation} \label{def_neel}
\underline{\sigma} = \{ \{ \sigma_j \} \,;\, j=1,...,N_p \, \,;\, \sigma_j = \left[ (j-1) \, \mbox{mod} \, N \right]  + 1\} \ , 
\end{equation}
which we call generalized Neel order for SU($N$) particles. Sectors with desired spin configurations can be accessed in experiments with ultracold atoms in optical lattices via spin-resolved manipulation techniques \cite{weitenberg_11, hild_fukuhara_14}.

\subsection{Spin-charge decomposition} \label{sec2_2}

Our main goal is to calculate nonequilibrium properties of spin-resolved one-body correlations $C^\sigma_l(x;t)$ between sites $l$ and $l+x$
\begin{align} \label{C_t_def}
C^\sigma_l(x;t)& =\langle\Psi(t)| \hat f^{(\sigma)\dagger}_{l+x} \hat f^{(\sigma)}_l |\Psi(t)\rangle \, ,
\end{align}
where $|\Psi(t)\rangle$ is the time-evolving state of the DQP model. We are also interested in computing the total one-body correlations
\begin{equation} \label{def_C_integrated}
C_l(x;t)= \sum_{\sigma=1}^{N} C_l^{\sigma}(x;t) \, ,
\end{equation}
which contain the contributions from all spin flavors.

Central to our approach is exploiting the separation of spin and charge degrees of freedom at infinite onsite repulsion. In particular, we use a compact representation of charge degrees of freedom as spinless fermions and spin degrees of freedom as nonlocal string of operators (projectors) expressed in terms of spinless fermions. Formal steps demonstrating the existence of such representations were reported in Ref.~\cite{kumar_09}. Here we use an explicit representation of the spin and charge decomposition for states that exhibit a generalized Neel order [see Eq.~\eqref{def_neel}]. This representation was introduced in Ref.~\cite{zhang18} and allows one to efficiently (in polynomial time) calculate spin-resolved one-body correlations numerically. 

Our approach is not based on the Bethe ansatz solution of the infinitely repulsive Fermi-Hubbard model~\cite{lieb_wu_68, ogata_shiba_90, parola_sorella_90, parola_sorella_92, penc_96, penc_hallberg_97, izergin_pronko_98, essler-book}. It has analogies with the Jordan-Wigner transformation used to map spin-$\frac{1}{2}$ systems onto spinless fermions. Since the pioneering work by Lieb, Schultz, and Mattis~\cite{lieb61}, the Jordan-Wigner transformation has been used as a standard tool to study quantum magnetism. It can be efficiently implemented numerically using properties of Slater determinants~\cite{rigol05a}, and has been used to study quantum quenches in 1D hard-core boson systems~\cite{rigol04, rigol05, rigol05a, rigol_dunjko_07, rigol_muramatsu_06, vidmar15, vidmar_iyer_17, xu_rigol_17, vidmar_xu_17}.

Using a compact representation of the spin-charge decomposition of states with a generalized Neel order~\cite{zhang18}, we rewrite Eq.~(\ref{C_t_def}) as
\begin{equation} \label{C_t_sf}
C^\sigma_l(x;t) = \langle\Psi_{\rm SF}(t)| \hat c^{\dagger}_{l+x} \hat c^{}_l\, {\cal \hat P}^{(\sigma)}_{l,x} |\Psi_{\rm SF}(t)\rangle \,.
\end{equation}
Here, $|\Psi_{\rm SF}(t)\rangle$ is the wave function that describes the charges (spinless fermions) at time $t$, while the time-independent projection operator ${\cal \hat P}^{(\sigma)}_{l,x}$ properly accounts for the spins. The Hamiltonian that governs the dynamics of the charge degrees of freedom $|\Psi_{\rm SF}(t)\rangle$ is the spinless fermion Hamiltonian
\begin{equation}\label{def_Hsf}
 \hat H_{\rm SF} = -\sum_{l=1}^{L-1}{(\hat c^{\dagger}_l \hat c^{}_{l+1}+\hat c^{\dagger}_{l+1} \hat c^{}_l)}\,,
\end{equation}
where $\hat c^\dagger_l$ ($\hat c^{}_l$) is the spinless fermion creation (annihilation) operator at site $l$.

The spin projection operator ${\cal \hat P}^{(\sigma)}_{l,x}$ is in general a nonlocal multi-body operator~\cite{zhang18}. It is constructed as the product of two operators
\begin{equation} \label{def_projP}
{\cal \hat P}^{(\sigma)}_{l,x} = {\cal \hat M}_{l,x}{\cal \hat R}^{(\sigma)}_{l}\,.
\end{equation}
The role of the operator ${\cal \hat M}_{l,x}$ is to prevent exchange of fermions. It is defined as
\begin{equation} \label{def_projM}
 {\cal \hat M}_{l,x}=\prod_{j=l+1}^{l+x-1} \left(1- \hat c_j^\dagger \hat c^{}_j \right) \,,
\end{equation}
and guarantees that all lattice sites from ${l+1}$ through $l+x-1$ are empty. The role of ${\cal \hat R}^{(\sigma)}_{l}$ is to target spin flavor $\sigma$ at site $l$. For spin configurations that have generalized Neel order [Eq.~(\ref{def_neel})], ${\cal \hat R}^{(\sigma)}_{l}(N)$ is defined as
\begin{equation} \label{def_projR}
{\cal \hat R}^{(\sigma)}_{l}(N) = \frac{1}{N} \sum_{k=0}^{N-1}{e^{-\frac{2\pi i}{N}\sigma k}\exp\left[\frac{2\pi i}{N}k\sum_{j=1}^l \hat c_j^\dagger \hat c^{}_j \right]} \, .
\end{equation}
This operator counts the number of fermions from sites $1$ through $l$, and ensures that it is consistent with  a fermion with spin flavor $\sigma$ occupying site $l$.

For the total one-body correlations $C_l(x;t)$, the spin projection operators can be further simplified. In analogy to equilibrium calculations~\cite{zhang18}, we get
\begin{equation} \label{avg_corr}
C_l(x;t) = \langle\Psi_{\rm SF}(t)| \hat c^{\dagger}_{l+x} \hat c_l^{}\, {\cal \hat M}_{l,x} |\Psi_{\rm SF}(t)\rangle \, .
\end{equation}
$C_l(x;t)$ is hence independent of the number of spin flavors $N$. Equation~\eqref{avg_corr} is valid for any spin orderings $\underline{\sigma}$ that fulfill Eq.~(\ref{def_sigma_dqp}).

\subsection{Implementation for quantum quenches} \label{sec2_3}

To evaluate Eq.~\eqref{C_t_sf}, we follow the approach used for hard-core bosons in Refs.~\cite{rigol04, rigol05a}. We first note that, given an initial state whose charge degrees of freedom can be written as a Slater determinant $|\Psi_{\rm SF}^I\rangle = \prod_{j=1}^{N_p} \sum_{m=1}^L A_{mj} \hat c_m^\dagger |\emptyset\rangle$, where $A_{mj}$ are matrix elements of an $L \times N_p$ matrix ${\bf A}$, the charge degrees of freedom of the time-evolving state can be written as
\begin{equation}\label{time_evolve_state}
|\Psi_{\rm SF}(t)\rangle= e^{-i\hat{H}_{\rm{SF}}t} |\Psi_{\rm{SF}}^I\rangle=\prod_{j=1}^{N_p}\sum_{m=1}^L G_{mj}(t) \hat c^\dagger_m|\emptyset\rangle\, ,
\end{equation}
where $G_{mj}(t)$ are the matrix elements of an $L \times N_p$ matrix ${\bf G}(t) = {\bf U} e^{-i {\bf E}_{\rm SF} t} {\bf U}^\dagger {\bf A}$. Here, ${\bf E}_{\rm SF}$ is a diagonal matrix that satisfies ${\bf H}_{\rm SF} {\bf U} = {\bf U} {\bf E}_{\rm SF}$, and ${\bf H}_{\rm SF}$ is the single-particle matrix representation of $\hat H_{\rm SF}$.

A crucial next step is to evaluate ${\cal \hat P}^{(\sigma)}_{l,x} |\Psi_{\rm SF}(t)\rangle$. As in Ref.~\cite{zhang18}, the spin projection operator defined in Eqs.~(\ref{def_projP})--(\ref{def_projR}) acting on $|\Psi_{\rm SF}(t)\rangle$ yields a linear combination of Slater determinants,
\begin{equation}
{\cal \hat P}^{(\sigma)}_{l,x}|\Psi_{\rm SF}(t)\rangle=\frac{1}{N}\sum_{k=0}^{N-1} e^{-\frac{2\pi i}{N}\sigma k}\prod_{j=1}^{N_p}\sum_{m=1}^L G^k_{mj}(t) \hat c^\dagger_m|\emptyset\rangle \,,
\end{equation}
where
\begin{equation} \label{def_Gijk}
G^k_{mj}(t)= \Bigg \{
	\begin{tabular} {ccl}
	$e^{\frac{2\pi i}{N}k}\,G_{mj}(t)$,&&$m\leq l$ \\
	$0$, && $l<m<l+x$ \\
        $G_{mj}(t)$, && ${\rm otherwise}$
	\end{tabular} \, 
\end{equation}
are matrix elements of a $L \times N_p$ matrix ${\bf G}^k(t)$.

The next to last step to evaluate Eq.~(\ref{C_t_sf}) is to act with the spinless fermion annihilation and creation operators on ${\cal \hat P}^{(\sigma)}_{l,x}|\Psi_{\rm SF}(t)\rangle$. We first rewrite Eq.~(\ref{C_t_sf}) as $C^\sigma_l(x;t)=\delta_{x,0} \langle\Psi_{\rm SF}(t)|{\cal \hat P}^{(\sigma)}_{l,x}|\Psi_{\rm SF}(t)\rangle-\langle\Psi_{\rm SF}(t)| \hat c^{}_{l} \hat c^{\dagger}_{l+x} {\cal \hat P}^{(\sigma)}_{l,x} |\Psi_{\rm SF}(t)\rangle$. In the second term, $\hat c^\dagger_{j} |\Psi_{\rm SF}(t)\rangle$ and $\hat c^{\dagger}_{j} {\cal \hat P}^{(\sigma)}_{l,x} |\Psi_{\rm SF}(t)\rangle$ result in adding an extra column to the matrices representing the Slater determinants involved, yielding ${\bf G}(t) \to {\bf G}'(t;j)$ and ${\bf G}^k(t) \to {\bf G}'^{k}(t;j)$, respectively, where $G_{l,N_p+1}' = G_{l,N_p+1}'^{k} = \delta_{l,j}$ ($G_{li}' = G_{li}^{}$ and $G_{li}'^{k} = G_{li}^{k}$, for $i \leq N_p$). Finally, computing the inner product of the resulting Slater determinants, one obtains
\begin{align} \label{C_t_Slater}
C^\sigma_l(x;t)= \frac{1}{N}\sum_{k=0}^{N-1}e^{-\frac{2\pi i}{N}\sigma k}
&\left( \delta_{x,0}\det[{\bf G}(t)^\dagger {\bf G}^k(t)]\right.\\ &\left.-\det[{\bf G}'(t;l)^\dagger {\bf G}'^k(t;l+x)] \right) \,.\nonumber
\end{align}
Note that, for time-evolving states after quantum quenches, Eq.~(\ref{C_t_Slater}) is equivalent to Eq.~(19) in Ref.~\cite{zhang18}.

We are particularly interested in the total one-body correlation function $C_l(x;t)$ defined in Eq.~(\ref{def_C_integrated}). It was shown in Ref.~\cite{zhang18} that $C_l(x)$ in equilibrium exhibits weak finite-size effects and allows one to learn about universal properties of DQPs that are independent of the specific spin configuration $\underline{\sigma}$. $C_l(x;t)$ can be computed as
\begin{align}\label{C_l}
C_l(x;t) =&\, \delta_{x,0}\det[{\bf G}(t)^\dagger {\bf G}^{k=0}(t)] \nonumber \\ 
&- \det[{\bf G}'(t;l)^\dagger {\bf G}'^{k=0}(t;l+x)] \,,
\end{align}
where the matrix elements of ${\bf G}^{k=0}(t)$ are given by $G_{mj}^k(t)$ in Eq.~(\ref{def_Gijk}) for $k=0$.

Before moving on to applications, we comment on a subtlety regarding correlations that are summed over all spin flavors. The total site occupations $n(l;t) = C_l(0;t)$ are identical to those of the noninteracting spinless fermions onto which we map the impenetrable SU($N$) fermions. On the other hand, for $x>1$, the total one-body correlations $C_l(x;t)$ are in general different from those of the spinless fermions. In fact, in Ref.~\cite{zhang18} it was shown that $C_l(x)$ exhibits a Gaussian decay in the ground state of the DQP model, while it exhibits a power-law decay in the ground state of spinless fermions.

\section{Dynamical emergence of Gaussian one-body correlations} \label{sec3}

As a first application of the approach introduced in the previous section, we study the expansion of (generalized) Neel domain walls. The initial state $|\Psi^I\rangle$ for the Fermi-Hubbard model ($N=2$) is shown in Fig.~\ref{fig1}(a). It is a product state with particles occupying the $N_p$ leftmost lattice sites,
\begin{equation}\label{init_state}
|\Psi^I\rangle=\prod_{j=1}^{N_p}\hat{f}^{(\sigma_j)\dagger}_j|\emptyset\rangle\, ,
\end{equation}
and the spin flavors exhibit the generalized N\'eel order defined in Eq.~(\ref{def_neel}). A computational implementation of Eqs.~(\ref{C_t_Slater}) and~(\ref{C_l}) allows us to straightforwardly solve systems with of the order of $L\sim10^3$ sites. We set $L=2N_p$ in our numerical calculations and only consider times at which particles moving to the right (or holes moving to the left) have not yet reached the lattice boundaries, namely, times $t < N_p/v_{\rm max}$, where $v_{\rm max} = 2$ is the maximal group velocity in the lattice. 

As $N_p\rightarrow\infty$, our setup is equivalent to the expansion of a semi-infinite domain wall in an infinite lattice. Such an expansion gives rise to a current-carrying steady state. In lattice Hamiltonians without internal degrees of freedom, properties of the current-carrying states emerging from an initial domain wall, as defined in Eq.~(\ref{init_state}), have been studied for spinless fermions~\cite{antal99}, hard-core bosons~\cite{rigol04}, and for the anisotropic Heisenberg model~\cite{gobert05}.

\begin{figure}[!t]
\begin{center}
\includegraphics[width=0.99\columnwidth]{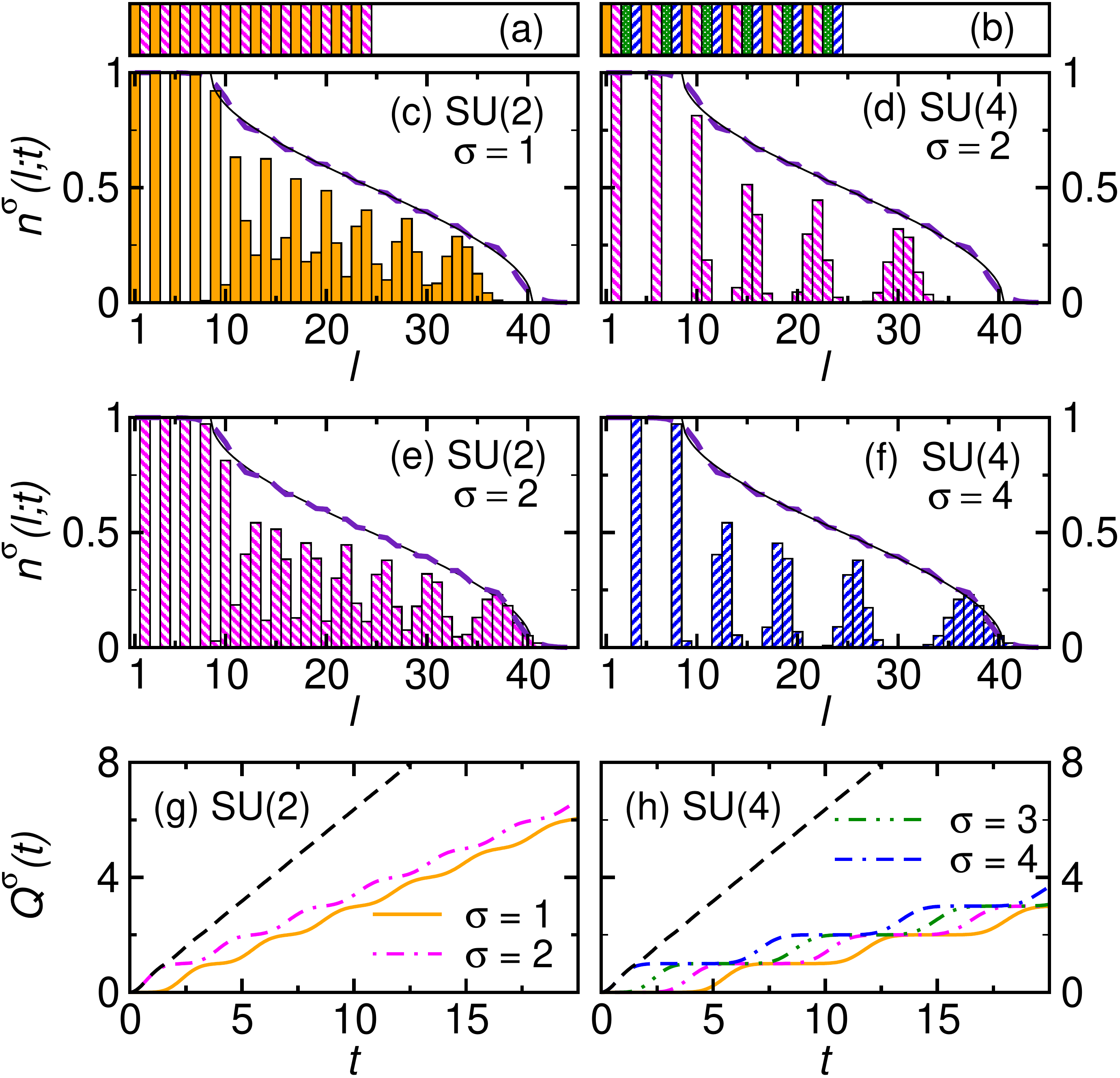}
\caption{Spin resolved site occupations $n^{\sigma}(l;t)$ during the ``melting'' of a domain wall~[see Eq.~(\ref{init_state})]. (a) [(b)] Initial SU(2) [SU(4)] domain walls with generalized Neel order (the spin flavors are encoded in the patterns and colors used). (c) and (e) [(d) and (f)] $n^{\sigma}(l;t)$ for $\sigma=1$ and $\sigma=2$ in the SU(2) case [$\sigma=2$ and $\sigma=4$ in the SU(4) case], respectively, at $t=8$. Thick dashed lines in (c)--(f) show the total site occupations $n(l;t)$, while the overlapping thin solid lines show the analytical predictions $n(l;t)_\infty$ from Eq.~(\ref{sf_siteoccup}) using $l_0 = N_p + 1/2$. (g) [(h)] Integrated number of particles $Q^\sigma(t) = \sum_{l > l_0} n^\sigma(l;t)$ that have crossed the initial edge of the domain wall (at $l_0$) for the SU(2) [SU(4)] case. Black dashed lines depict results for spinless fermions $\sum_\sigma Q^\sigma(t)$. Numerical results are shown for $N_p=24$ particles in (a)--(f) and for $N_p=120$ particles in (g) and (h).}\label{fig2}
\end{center}
\end{figure}

Figures~\ref{fig2}(a) and~\ref{fig2}(b) show $n^{\sigma}(l;t=0)$ in the initial generalized Neel domain wall for  SU(2) and SU(4) systems, respectively, with 24 fermions. Figures~\ref{fig2}(c)--\ref{fig2}(f) show the corresponding $n^{\sigma}(l;t)$ at time $t=8$. The latter make apparent an interesting feature of the domain-wall melting in our setup. At short times, the spin-resolved site occupations $n^{\sigma}(l;t) \equiv \langle\Psi(t)|\hat{f}^{(\sigma)\dagger}_l \hat{f}^{(\sigma)}_l |\Psi(t)\rangle$ resolve individual quantum particles quite well. The ``resolution'' increases with increasing $N$, as made apparent by comparing $N=2$ and $4$. Figures~\ref{fig2}(c)--\ref{fig2}(f) also show the total site occupations $n(l;t) = \sum^{N}_{\sigma=1} n^{\sigma}(l;t) = C_l(0;t)$ (dashed lines), which are identical to those of the corresponding spinless fermions.

As shown in an exact calculation~\cite{antal99}, as well as using hydrodynamic arguments~\cite{antal08}, $n(l;t)$  during the domain-wall melting obeys the scaling form
\begin{equation} \label{sf_siteoccup}
n(l;t)_\infty=\frac{1}{\pi}\arccos \left( \frac{l-l_0}{2t} \right) \, ,
\end{equation}
when both $l, t \to \infty$, while their ratio is kept fixed within the region $-1\leq(l-l_0)/(2t)\leq1$. Here, $l_0$ denotes the initial domain-wall edge. Unless stated otherwise, in this section we set $l_0=N_p$. Outside the scaling region, site occupations simply equal $n(l;t) = 1$ (to the left) and $n(l;t) = 0$ (to the right). The thin solid lines in Figs.~\ref{fig2}(c)--\ref{fig2}(f) show $n(l;t)_\infty$. They fit very well the numerical results in finite systems. We identify the current-carrying steady state as that in the scaling regime, namely, as that in which local properties only depend on the ratio $(l-l_0)/t$.

Figures~\ref{fig2}(g) and~\ref{fig2}(h) show the integrated number of particles of a given spin flavor $\sigma$ that have crossed $l_0$ ($l_0=N_p+1/2$ in Fig.~\ref{fig2}) at time $t$, defined as $Q^\sigma(t) = \sum_{l > l_0} n^\sigma(l;t)$. $Q^\sigma(t)$ exhibits clear plateaus at integer values (better seen for $N=4$). Those values increase by one each time that a particle from the domain wall (with a given $\sigma$) crosses the initial edge. In contrast, a steady linear growth is seen for spinless fermions [defined as $\sum_\sigma Q^\sigma(t)$, shown as dashed lines in Figs.~\ref{fig2}(g) and~\ref{fig2}(h)] reflecting the ballistic nature of the charge current in the system.

\subsection{Emergent eigenstate solution}

One of the most remarkable features of the current-carrying states generated by the melting of domain walls such as the ones in Fig.~\ref{fig2}, observed in studies involving spinless fermions~\cite{antal99, eisler09}, hard-core bosons~\cite{rigol04, rigol05a}, Bose-Hubbard~\cite{rodriguez_manmana_06} and Fermi-Hubbard~\cite{hm08} models, and Bose gases~\cite{denardis_panfil_18}, is the dynamical emergence of one-body correlations with ground-state-like behavior. We explore one-body correlations in our time-evolving states to see whether they exhibit a Gaussian decay like the one observed in the ground state of impenetrable SU($N$) fermions constrained to the same spin configurations~\cite{zhang18}.

A theoretical framework that provides an understanding of the dynamical emergence of ground-state-like correlations during expansion dynamics was put forward in Ref.~\cite{vidmar_iyer_17}. It was dubbed {\it emergent eigenstate solution} to quantum dynamics as it allows one to construct a local operator (called the {\it emergent Hamiltonian}) of which the time-evolving state is an eigenstate. For the initial domain wall of spinless fermions, the emergent local Hamiltonian is (up to a constant)
\begin{equation}\label{emergent_sharpwall}
\hat{\cal{H}}_{\rm{SF}}(t) = -t \sum_{l=1}^{L-1}(i\hat c^{\dagger}_l \hat c^{}_{l+1}+ \mathrm{H.c.})+\sum_{l=1}^L l\hat{n}_l\,,
\end{equation}
where $\hat{n}_l=\hat c^{\dagger}_l \hat c_l$ is the particle-number operator at site $l$~\cite{vidmar_iyer_17}. Note that the expansion time $t$ is a parameter in $\hat{\cal{H}}_{\rm{SF}}(t)$. The time-evolving state is the ground state of $\hat{\cal{H}}_{\rm{SF}}(t)$ (up to corrections that vanish exponentially with system size) as long as particles (holes) do not reach the lattice boundary, i.e., as long as $n(1;t)=1$ and $n(L;t)=0$~\cite{vidmar_iyer_17}. Hence, in our setup, the ground state of $\hat{\cal{H}}_{\rm{SF}}(t)$ in Eq.~(\ref{emergent_sharpwall}), denoted by $|\Phi_{\rm{SF}}(t)\rangle$, provides an accurate (exact in the limit  $N_p\rightarrow\infty$) description of the time-evolving state for the charge degrees of freedom of the impenetrable SU($N$) fermions $|\Psi_{\rm SF}(t)\rangle$. We then calculate one-body correlations $C^\sigma_l(x;t)$ via $|\Phi_{\rm{SF}}(t)\rangle$ using the spin projectors introduced in Sec.~\ref{sec2_2}.

While in previous applications of the emergent eigenstate solution $|\Phi_{\rm{SF}}(t)\rangle$ was evaluated exactly numerically~\cite{vidmar_iyer_17, vidmar_xu_17, modak_vidmar_17}, here we use the local density approximation (LDA) in the scaling regime to simplify $\hat{\cal{H}}_{\rm{SF}}(t)$. This approach is similar to the generalized hydrodynamic approach~\cite{bertini16, castroalvaredo_doyon_16}, in which a GGE is constructed for each rescaled position of the current-carrying state. We introduce, for $t>0$, an effective local chemical potential at the (continuous) rescaled position $s=(l-l_0)/t$,
\begin{equation}
\mu(s) = \mu_0 - s\, ,
\end{equation}
where $\mu_0$ is the global chemical potential. For each $s$, the system is homogeneous with an effective chemical potential $\mu(s)$:
\begin{equation} \label{Heme_lda}
\hat{\cal H}_{\rm SF}^{\rm LDA}(s) = - \sum_{l}(i\hat c^{\dagger}_l \hat c^{}_{l+1}+ \mathrm{H.c.}) - \mu(s) \hat N \, .
\end{equation}

We test the emergent eigenstate solution plus LDA by analytically calculating the site occupations $n(l;t)$ for spinless fermions. They follow $n(s)=\arccos[-\mu(s)/2]/\pi$ when $-2 <\mu(s) < 2$, and $n(s)=0$ [$n(s)=1$] for $\mu(s)\leq-2$ [$\mu(s)\geq 2$]. We fix $\mu_0$ using the total particle number, $N_p=\int n(s) ds$, which yields the final expression for site occupations
\begin{equation}
n(s)=\frac{1}{\pi}\arccos\left( \frac{s}{2} \right) ,
\end{equation}
when $-1\leq s/2 \leq1$. This is the result in Eq.~(\ref{sf_siteoccup}).

\subsection{Total one-body correlations}

\begin{figure}[!t]
\begin{center}
\includegraphics[width=0.99\columnwidth]{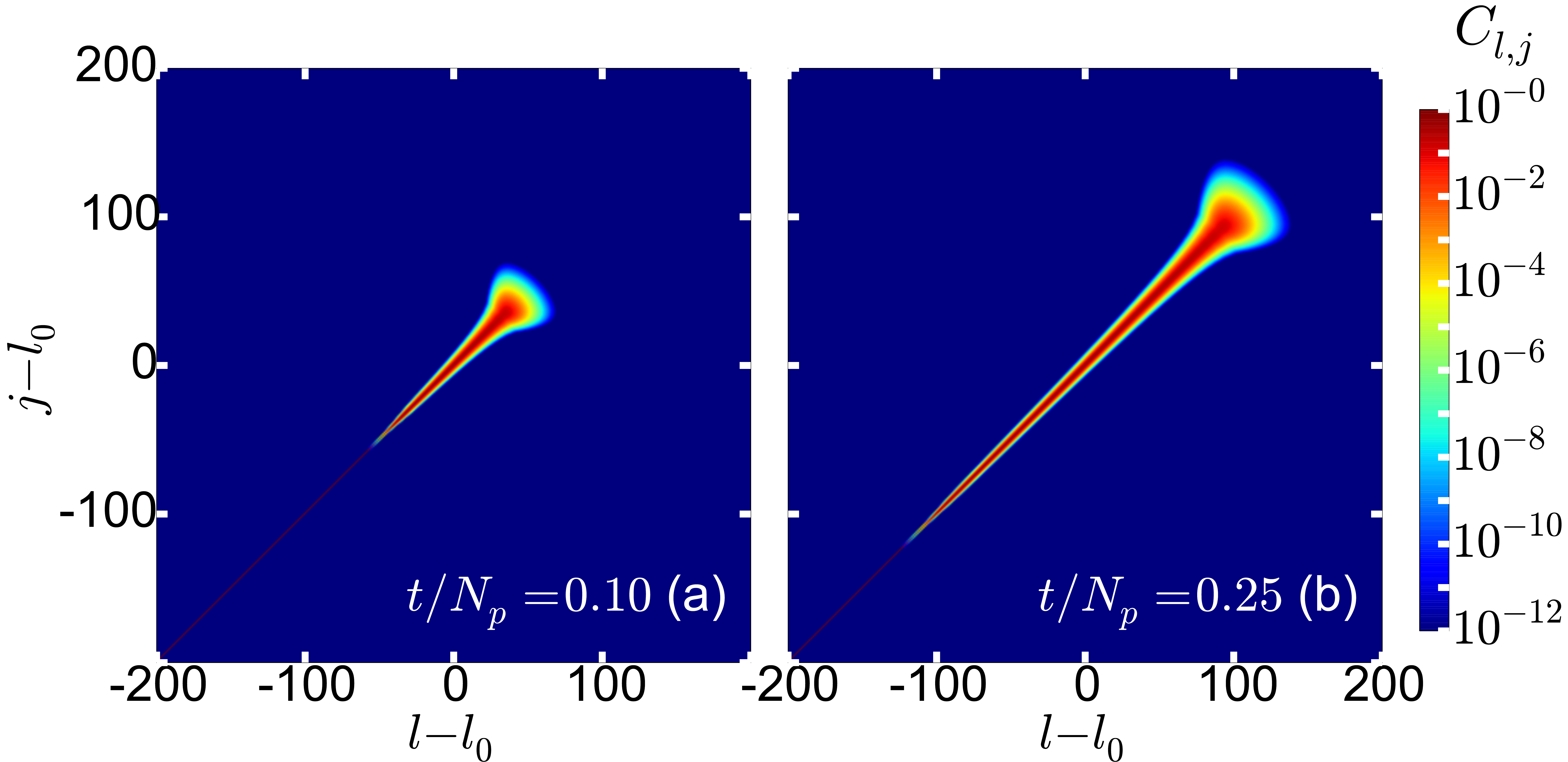}
\caption{Matrix elements $C_{l,j} = |C_l(j-l;t)|$ of the total one-body correlation matrix for an initial domain wall with $N_p=200$ particles [see Eq.~(\ref{init_state})]. Results are shown at times (a) $t/N_p=0.1$ and (b) $t/N_p=0.25$. $l_0=N_p$ is the initial edge of the domain wall.}\label{fig3}
\end{center}
\end{figure}

\begin{figure}[!t]
\begin{center}
\includegraphics[width=0.99\columnwidth]{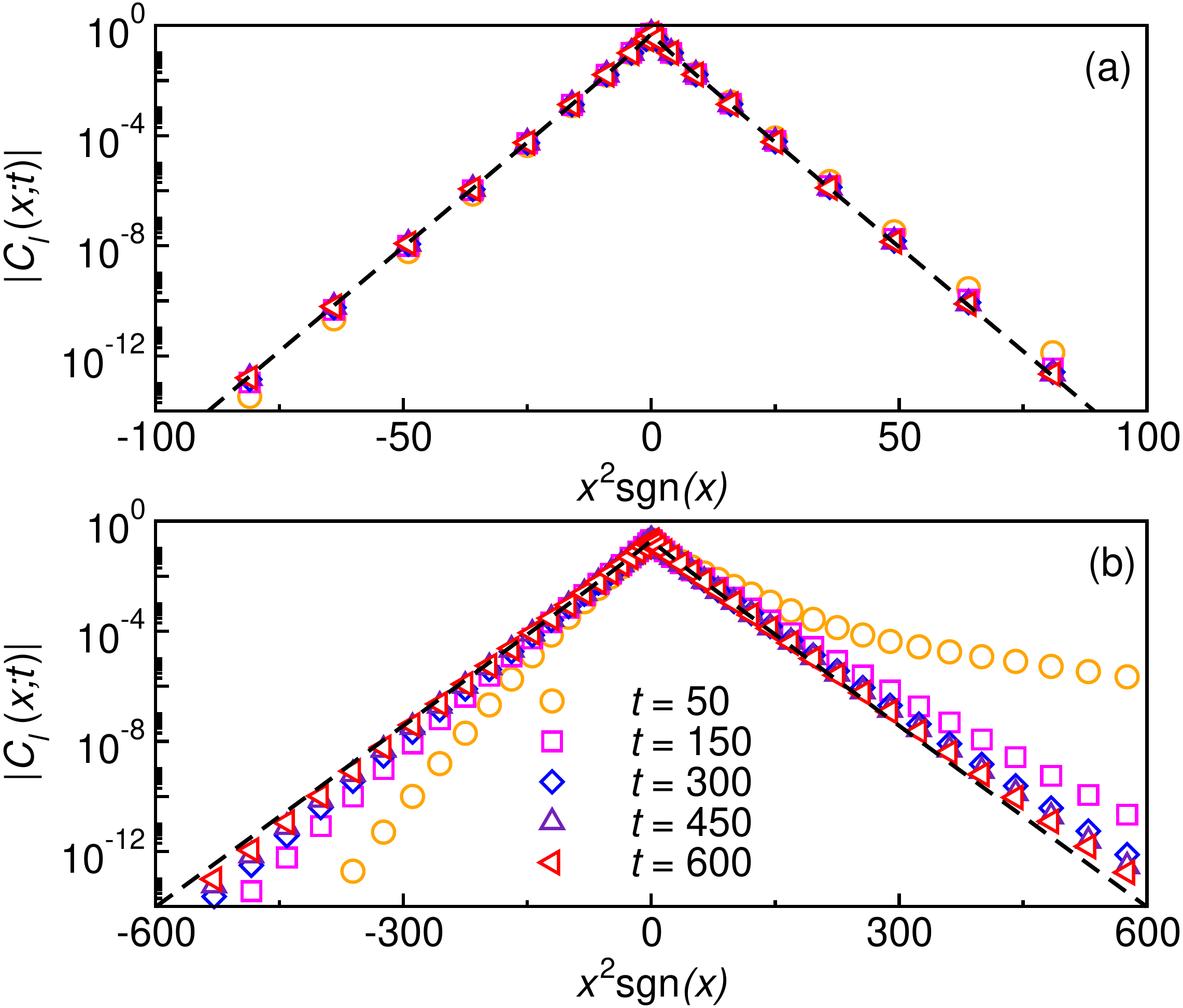}
\caption{Total one-body correlation function $|C_l(x;t)|$ for an initial domain wall with $N_p=1200$ particles [see Eq.~(\ref{init_state})]. We set the target particle occupation $n(l;t) = n_0$ and use Eq.~(\ref{sf_siteoccup}) to find the optimal $l$ at a given time $t$. (a) $n_0=0.5$, which corresponds to $l=1200$ at all times. (b) $n_0=0.2$, which corresponds to $l=1281,\, 1443,\, 1685,\, 1928,\, 2171$ at times $t=50,\, 150,\, 300,\, 450,\, 600$, respectively. Dashed lines are results for $C_l(x)$ in the ground state of the emergent local Hamiltonian $\hat{\cal H}_{\rm SF}^{\rm LDA}(s)$ [see Eq.~(\ref{Heme_lda})]. For the latter calculations, we set the particle filling $N_p/L=0.5$ in (a) and $N_p/L=0.2$ in (b), and measure correlations from the center of the lattice ($l=L/2$, $L=2400$).}\label{fig4}
\end{center}
\end{figure}

Next, we study the off-diagonal one-body correlations in the current-carrying state. They are exactly zero in the initial product state [Eq.~(\ref{init_state})], and become nonzero as a result of the expansion dynamics. To minimize finite-size effects (see Appendix~\ref{app1}), we focus on the total correlations $C_l(x;t)$ [see Eq.~(\ref{C_l})]. As an example, Fig.~\ref{fig3} shows the absolute value of all matrix elements of the correlation matrix $C_{l,j} = |C_l(j-l;t)|$ at times $t=20$ and $50$, in a system with $N_p=200$. Those results reveal a very fast decay of the one-body correlations with $x = j-l$, in stark contrast with the power-law decay observed for spinless fermions (see, e.g., Fig.~2 in Ref.~\cite{vidmar_iyer_17}).

Figure~\ref{fig4} shows one-body correlations vs $x$ measured with respect to sites with a given particle occupation $n(l;t) = n_0$, for a system larger than the one in Fig.~\ref{fig3}. We answer three questions about the current-carrying steady state:\\
(i) Does $|C_l(x;t)|$ exhibit a Gaussian decay with $x$ for different values of $l$ and $t$?\\
(ii) Do off-diagonal one-body correlations obey the scaling solution
\begin{equation}
 |C_l(x;t)| \stackrel{?}{=} |C_s(x)| \;.
\end{equation}
(iii) Do the exact numerical results for $|C_l(x;t)|$ match the predictions from the emergent eigenstate solution plus LDA?

Figure~\ref{fig4}(a) shows one-body correlations with respect to a site with $n_0=0.5$ at different times [see legends in Fig.~\ref{fig4}(b)]. It is apparent that one-body correlations exhibit a Gaussian decay that is independent of time. They only depend on the filling of the reference site, equivalently, $|C_l(x;t)| = |C_s(x)|$. The same conclusion applies, at long times, to the results for $n_0=0.2$ shown in Fig.~\ref{fig4}(b). They converge to a Gaussian decay that is independent of time, i.e., $|C_l(x;t)| = |C_s(x)|$ at long times. The onset of the scaling behavior for the off-diagonal one-body correlations occurs at later times as the reference site occupation departs from half-filling.

In Figs.~\ref{fig4}(a) and~\ref{fig4}(b), we show (as dashed lines) numerical results for one-body correlations in the ground state of the emergent local Hamiltonian [Eq.~(\ref{Heme_lda})] at the target particle filling. Those correlations are in excellent agreement with the ones in the current-carrying state at long times. We should add that the absolute value of the one-body correlations in the ground state of the emergent local Hamiltonian are identical to the ground-state correlations in the physical Hamiltonian that governs the dynamics, Eq.~(\ref{def_H})~\cite{vidmar_iyer_17}. The latter is known to exhibit one-body correlations with a Gaussian decay
\begin{equation}
\label{gaussian_peak} 
|C_s(x)| = n(s) \, e^{-x^2/x_0^2}\,,
\end{equation}
where $x_0\propto n(s)^{-1}$~\cite{zhang18}.

\subsection{Quasimomentum distribution function}

\begin{figure}[!b]
\begin{center}
\includegraphics[width=0.99\columnwidth]{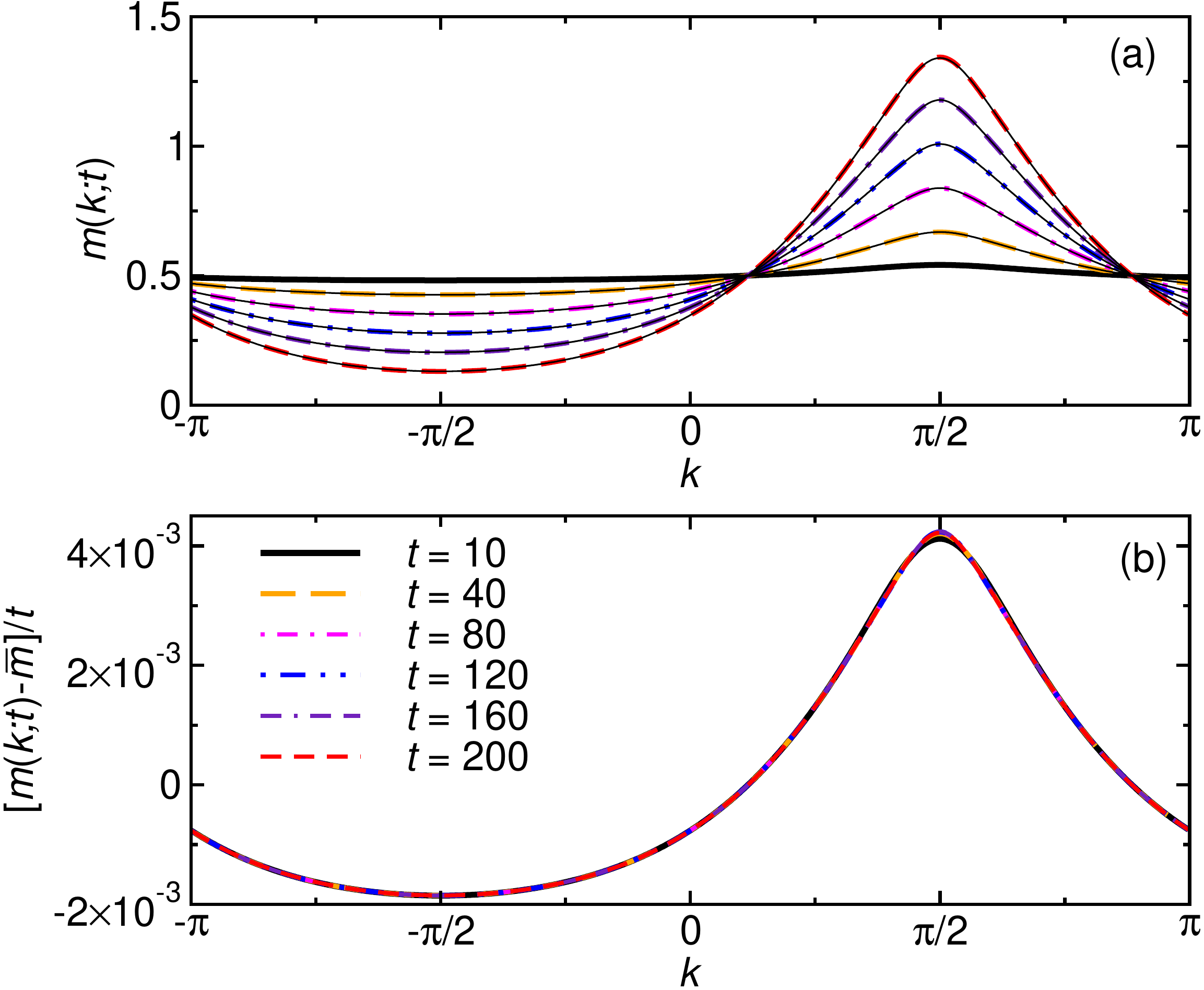}
\caption{Total quasimomentum distribution $m(k;t)$ during the melting of a domain wall [see Eq.~(\ref{init_state})] with $N_p=400$ particles. (a) $m(k;t)$ for different times. Thick lines are exact results for the dynamics while thin solid lines are results obtained in the ground state of the emergent local Hamiltonian $\hat {\cal H}(t)$ [see Eq.~(\ref{emergent_sharpwall})]. (b) Rescaled exact results for the quasimomentum distribution $[m(k;t)-\bar{m}]/t$, where $\bar{m}=N_p/L=0.5$.} \label{fig5}
\end{center}
\end{figure}

To conclude the study of the current-carrying state, we calculate the total quasimomentum distribution function 
\begin{equation}\label{eq:nk}
m(k;t)=\frac{1}{L}\sum_{l,x}e^{-ikx}C_l(x;t) \, ,
\end{equation}
which is routinely measured in experiments with ultracold quantum gases~\cite{bloch08}. (Spin-resolved results are shown in Appendix~\ref{app1}.)

Figure~{\ref{fig5}}(a) shows $m(k;t)$ at different times for a system with $N_p=400$ particles. In the initial product state, in which fermions are localized, all quasimomenta are equally occupied, i.e., $m(k;0)=\bar{m}=N_p/L=0.5$. Two main features are apparent in the dynamics: a peak emerges at $k=\pi/2$, and the height of the peak increases with time. Figure~\ref{fig5}(a) also makes apparent that the results obtained for $m(k;t)$ during the exact time evolution are indistinguishable from those obtained in the ground state of the emergent local Hamiltonian $\hat {\cal H}_{\rm SF}(t)$ [Eq.~(\ref{emergent_sharpwall})].

The emergence of the peak at $k=\pi/2$ can be understood from the phase $i=\exp(i\pi/2)$ in the hopping term of $\hat {\cal H}_{\rm SF}(t)$. As a result, the minimum of the energy dispersion of the kinetic part of $\hat {\cal H}_{\rm SF}(t)$ is located at $k=\pi/2$. For 1D hard-core bosons, which can also be mapped onto spinless fermions, the dynamical emergence of the peak at $k=\pi/2$~\cite{rigol04} has been experimentally observed in experiments with bosons in 1D optical lattices~\cite{vidmar15}.

For the current-carrying state of spinless fermions, it was shown analytically that the evolution of the  quasimomenta occupations relative to the initial occupations $m(k;t)-\bar{m}$ is linearly proportional to time $t$~\cite{vidmar_iyer_17}. In Fig.~\ref{fig5}(b), we plot the rescaled quasimomentum distribution $[m(k;t)-N_p/L]/t$ for the impenetrable SU($N$) fermions. We observe a perfect data collapse, confirming the expectation from Fig.~\ref{fig5}(a) that the height of the peak increases linearly with time.

\section{Unveiling the rapidity distribution} \label{sec4}

As a second application of the approach introduced in Sec.~\ref{sec2}, we study the expansion of harmonically trapped impenetrable SU($N$) fermions in a lattice after suddenly turning off the trap. We take the initial state to be the ground state of the infinitely repulsive (generalized) Fermi-Hubbard model~[Eq.~(\ref{def_H})] in the presence of a harmonic confining potential
\begin{equation}
\label{Hamiltonian_trap}
 \hat H'_N =\hat H_N+\frac{1}{R^2}\sum_{l=1}^{L}\sum_{\sigma=1}^{N}(l-l_0)^2\hat{f}_l^{\dagger(\sigma)}\hat{f}_l^{(\sigma)}\,,
\end{equation}
where $l_0=L/2$ is the trap center, and $R$ is the characteristic length of the trap. Since the observables studied in this section are summed over all spin flavors, our results apply equally to the two-component Fermi-Hubbard model ($N=2$) as they do to arbitrary $N$-flavor SU($N$) models, for initial states in which every pair of consecutive fermions carries distinct spin flavors [Eq.~(\ref{def_sigma_dqp})]. The corresponding spinless fermion Hamiltonian, to which we map the impenetrable SU($N$) fermions, has the form $\hat H'_{\rm SF} = \hat H_{\rm SF} + R^{-2}\sum_l (l-l_0)^2\hat{c}_l^{\dagger}\hat{c}^{}_l$.

At time $t=0$, the harmonic confinement is turned off (we set $R^{-1}=0$) and the system evolves under $\hat H_N$. We are interested in dynamics that occurs before particles reach the boundaries of the lattice [$n(1;t)=n(L;t)=0$], which are equivalent to dynamics in an infinite lattice.

\subsection{Dynamical emergence of rapidities} \label{sec4a}

\begin{figure}[!t]
\begin{center}
\includegraphics[width=0.99\columnwidth]{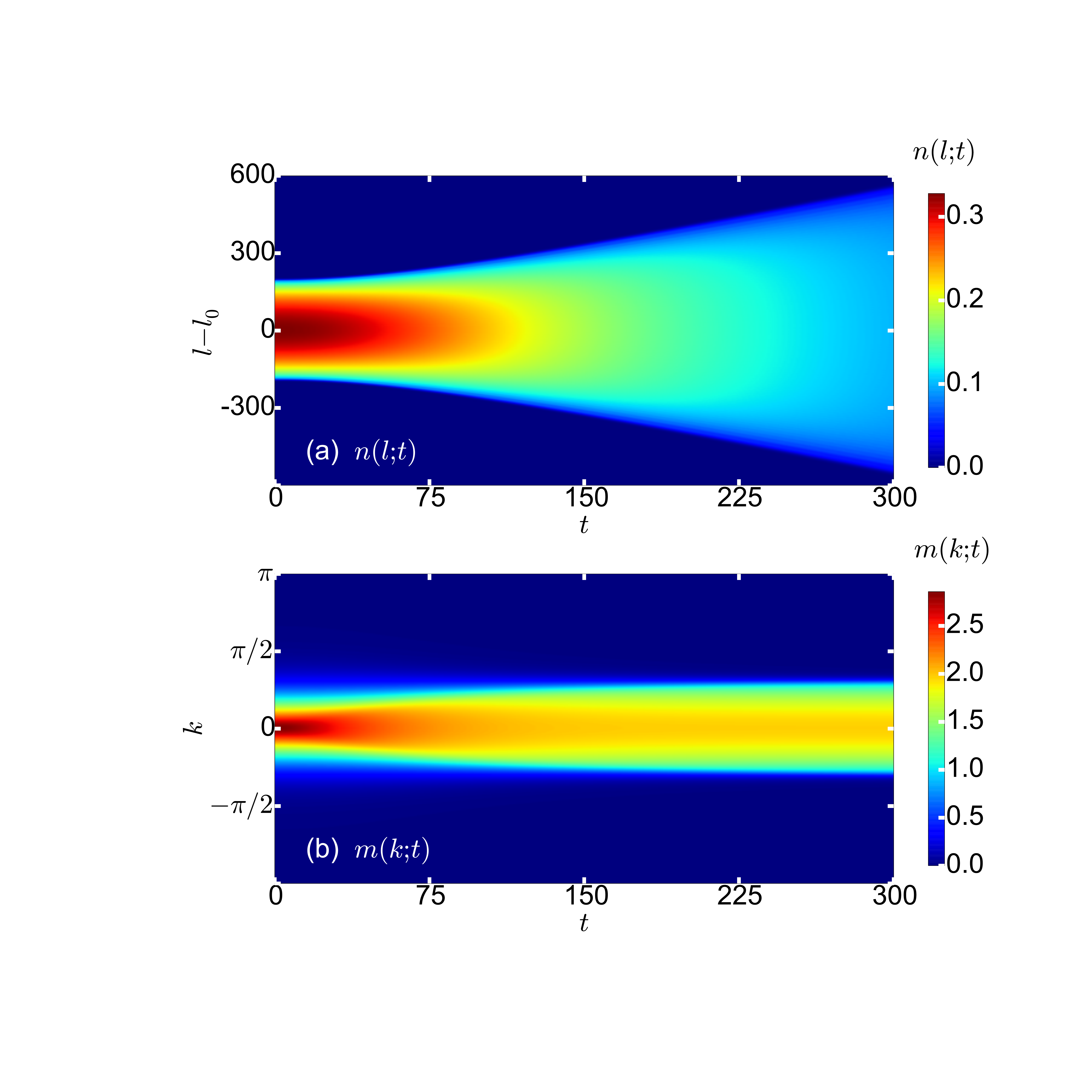}
\caption{Sudden expansion of $N_p=100$ impenetrable SU($N$) fermions after turning off a harmonic trap with $R=200$. We take $L=1200$ and $l_0=600$. (a) Total site occupations $n(l;t)$. (b) Total quasimomentum distribution $m(k;t)$.}\label{fig6}
\end{center}
\end{figure}

Figures~\ref{fig6}(a) and~\ref{fig6}(b) show density plots of the expanding total site occupation $n(l;t)$ and the quasimomentum distribution function $m(k;t)$, respectively, as functions of time. Figure~\ref{fig6}(b) shows that, after some transient time and despite the fact that $n(l;t)$ expands at all times, $m(k;t)$ becomes independent of time. [In this section, we normalize $m(k;t)$ using the characteristic length of the initial trap, $m(k;t)=\sum_{l,x}e^{-ikx}C_l(x;t)/R$.]

Our main focus next is to understand the stationary quasimomentum distribution function observed at long times. In various 1D bosonic systems including lattice hard-core bosons~\cite{rigol05, vidmar_xu_17}, as well as Tonks-Girardeau~\cite{minguzzi05} and Lieb-Liniger gases~\cite{buljan_pezer_08, jukic_klajn_09, iyer12, campbell15, piroli_calabrese_17}, it has been shown that during expansion dynamics the quasimomentum distribution function of the bosons undergoes a dynamical fermionization. For lattice hard-core bosons~\cite{rigol05, vidmar_xu_17} and Tonks-Girardeau gases~\cite{minguzzi05}, this fermionization is nothing but a transformation of the quasimomentum distribution function of the physical particles (impenetrable bosons) into the quasimomentum (rapidity) distribution of the underlying noninteracting fermions into which the former can be mapped. This transformation of quasimomentum distribution functions of physical particles into rapidity distributions under expansion dynamics is expected to be a generic phenomenon in integrable models~\cite{sutherland_98, bolech12, campbell15, mei16}.

\begin{figure}[!t]
\begin{center}
\includegraphics[width=0.99\columnwidth]{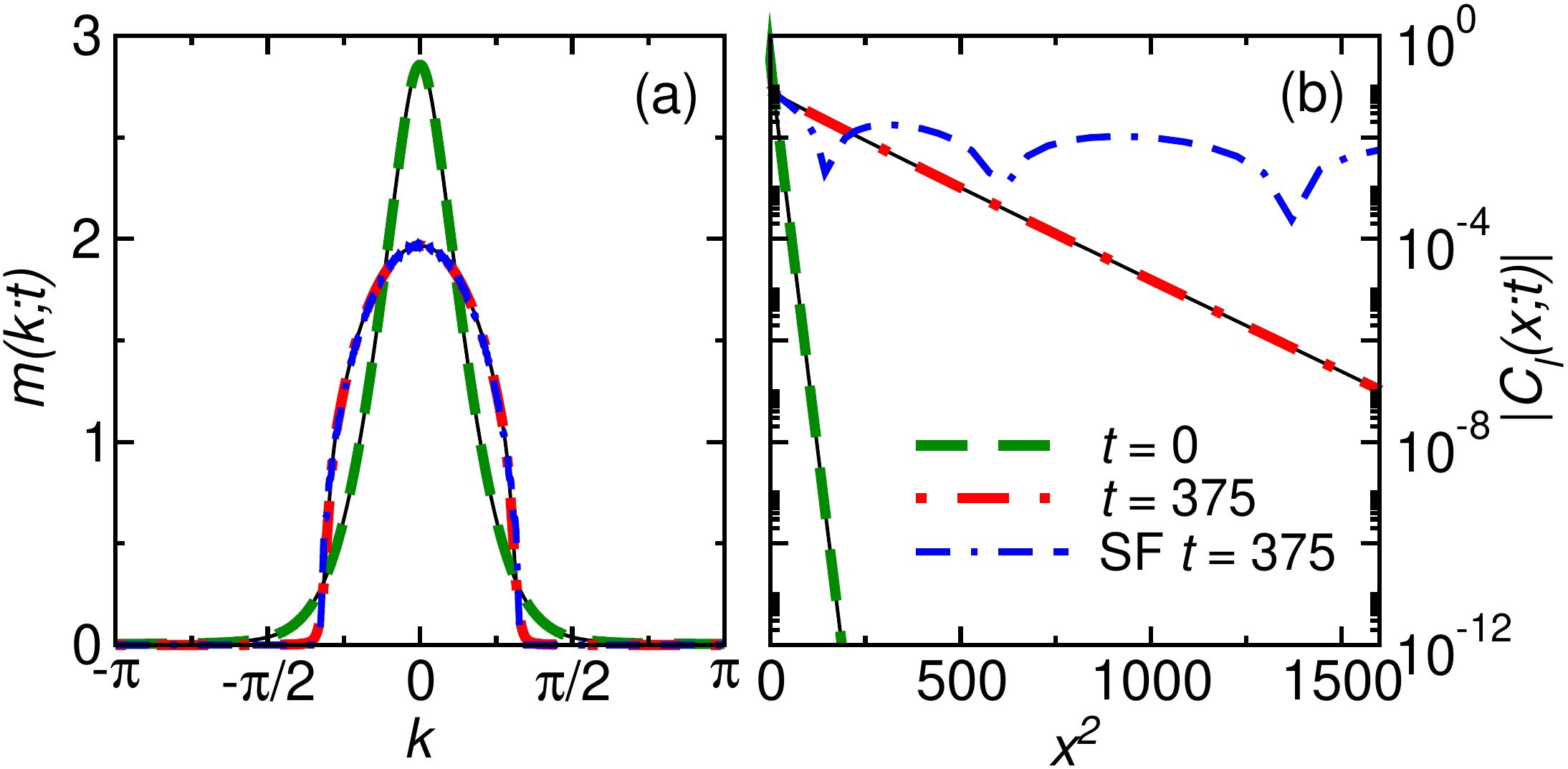}
\caption{Sudden expansion from a harmonic trap with $R=200$. Results are shown for impenetrable SU($N$) fermions in the initial state ($t=0$) and at $t=375$, and for spinless fermions (SF) at $t=375$. Thin solid lines depict the results for the impenetrable SU($N$) fermions in the ground state of the emergent local Hamiltonian~(\ref{def_Heme_2}). (a) Total quasimomentum distribution function $m(k;t)$. (b) Total one-body correlations $|C_l(x;t)|$ measured from the center of the system (at $l=1200$). The results shown are for $N_p=100$ and $L=2400$.}\label{fig7}
\end{center}
\end{figure}

Figure~\ref{fig7}(a) shows $m(k;t)$ in the initial state and at time $t=375$, a time at which $m(k;t)$ has already become stationary (corresponding spin-resolved results are shown in Appendix~\ref{app1}). We consider $N_p=100$ particles and an initial trap with $R=200$. Remarkably, the stationary result obtained for $m(k;t)$ is identical to the quasimomentum distribution function $m_{\rm SF}(k)$ of the spinless fermions to which we mapped the impenetrable SU($N$) fermions. $m_{\rm SF}(k)$ does not evolve in time because the quasimomenta occupations of the spinless fermions are conserved under expansion dynamics. $m_{\rm SF}(k)$ is the distribution of rapidities in our model, and our results show that it can be unveiled by allowing the impenetrable SU($N$) fermions freely expand up to about two to three times their initial size.

Figure~\ref{fig7}(b), on the other hand, shows that despite the agreement between the quasimomentum distribution of impenetrable SU($N$) fermions and the quasimomentum distribution function of spinless fermions, at long times, the off-diagonal one-body correlations of both systems are not the same (this also occurs for lattice hard-core bosons~\cite{rigol05, vidmar_xu_17} and Tonks-Girardeau gases~\cite{minguzzi05}). The absolute value of the one-body correlations exhibits a Gaussian decay for impenetrable SU($N$) fermions at all times and a power-law decay for spinless fermions. 

\subsection{Emergent eigenstate solution} \label{sec4b}

Figure~\ref{fig7}(b) also shows the one-body correlations of the impenetrable SU($N$) fermions at $t=0$. They exhibit a Gaussian decay as the ground state of homogeneous systems~\cite{zhang18}. The fact that in Fig.~\ref{fig7}(b) the decay of correlations has the same behavior in the ground state and during the expansion indicates that the time-evolving states share properties with the initial ground state. 

Like for the initial states considered in Sec.~\ref{sec3}, one can find an emergent eigenstate solution for the expansion of spinless fermions after turning off a harmonic trap~\cite{vidmar_xu_17}. The emergent local Hamiltonian for the corresponding spinless fermion system~\cite{vidmar_xu_17, modak_vidmar_17} is (up to a constant)
\begin{align} \label{def_Heme_2}
\mathcal{\hat {H}}_{\rm SF}'(t)=&-\sum_{l=1}^{L-1} A(l;t)(e^{i\phi(l;t)}\hat c_{l+1}^\dagger \hat c_{l}+\mathrm{H.c.})\\&-\frac{t^2}{R^2}\sum_{l=1}^{L-2}(\hat c^\dagger_{l+2}\hat c_l+\mathrm{H.c.})+\frac{1}{R^2}\sum_{l=1}^L(l-l_0)^2\hat c_{l}^\dagger\hat c_{l}\,,\nonumber
\end{align} 
where $A(l;t)=\sqrt{1+[(2t/R^2)(l-l_0+1/2)]^2}$ is the nearest-neighbor hopping amplitude, and the phase is $\phi(l;t)=\arctan[2t(l-l_0+1/2)/R^2]$. 

In Fig.~\ref{fig7}(b) we show (as thin solid lines) the correlations $|C_l(x;t)|$ obtained in the ground state of $\hat{\cal H}_{\rm SF}'(t)$, which perfectly overlap with the exact results from the nonequilibrium time evolution. Moreover, one can also combine the LDA with the emergent eigenstate solution, as in Sec.~\ref{sec3}, to describe the one-body correlations $|C_l(x;t)|$ in the nonequilibrium steady state in terms of those in a homogeneous system.

The appropriate homogeneous Hamiltonian can be obtained from $\mathcal{\hat {H}}_{\rm SF}'(t)$ replacing the harmonic trap by an effective local chemical potential, and making the amplitude $A(l;t)$ and the phase $\phi(l;t)$ site independent:
\begin{align}
\label{Emergent_LDA_trap}
\hat{\cal{H}}_{\rm SF}'^{\rm LDA}(t,l)=&-A(l;t)\sum_{j}[e^{i\phi(l;t)}\hat c_{j+1}^\dagger \hat c_{j}+\mathrm{H.c.}]\nonumber\\&-\frac{t^2}{R^2}\sum_{j}(\hat c_{j+2}^\dagger \hat c_{j}+\mathrm{H.c.}) + \mu(l) \hat N\,.
\end{align}

\begin{figure}[!tb]
\begin{center}
\includegraphics[width=0.99\columnwidth]{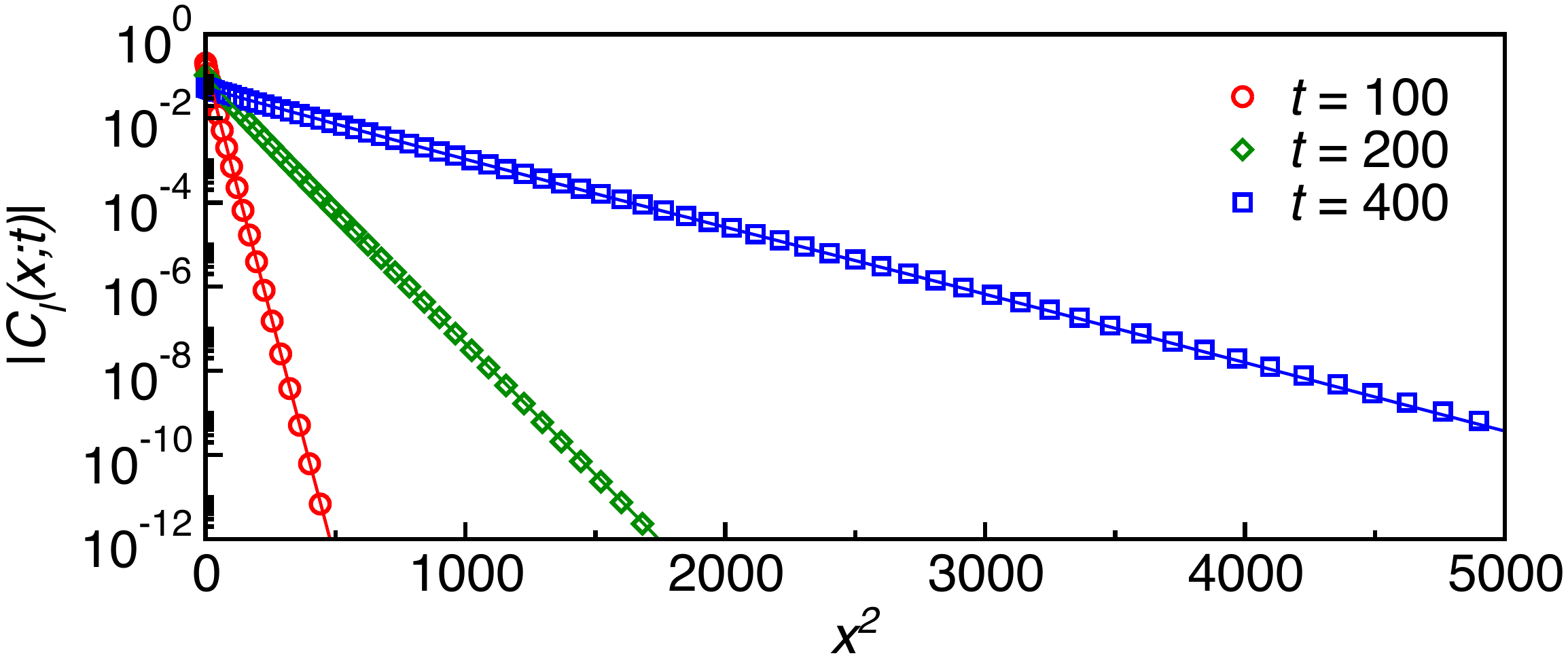}
\caption{Total one-body correlations $|C_l(x;t)|$ during the sudden expansion of $N_p=100$ impenetrable SU($N$) fermions after turning off a harmonic trap with $R=100$. We take $L=2400$ and measure correlations from the center of the system (at $l=1200$). The results from the exact time evolution are shown as symbols, and the results in the ground state of $\hat{\cal{H}}_{\rm SF}'^{\rm LDA}(t)$ [see Eq.~(\ref{Emergent_LDA_trap})] are shown as straight lines. The chemical potential in $\hat{\cal{H}}_{\rm SF}'^{\rm LDA}(t)$ is set to match the particle filling $n(l;t)$ at any given time.} \label{fig8}
\end{center}
\end{figure}

In Fig.~\ref{fig8} we compare $|C_l(x;t)|$ calculated during the exact time evolution (symbols) with the results in the ground state of $\hat{\cal{H}}_{\rm SF}'^{\rm LDA}(t)$ in Eq.~(\ref{Emergent_LDA_trap}), when the particle occupation matches the target site occupation $n(l;t)$ (lines). We show result for correlations measured from the trap center ($l=l_0=L/2$), for which $A(l;t) \approx 1$ and $\phi(l;t)\approx 0$, for systems with $N_p =100$ impenetrable SU($N$) fermions and an initial $R=100$. The results shown in Fig.~\ref{fig8} from both approaches yield a perfect agreement, and follow the functional form in Eq.~(\ref{gaussian_peak}).

\section{Equilibration and Generalized thermalization} \label{sec5}

As a third, and final, application of the approach introduced in Sec.~\ref{sec2}, we study the equilibration of impenetrable SU($N$) fermions in a finite lattice with open boundary conditions (equivalent to a box trap with infinitely high walls) after suddenly turning off a harmonic trap in which the impenetrable SU($N$) fermions are initially confined. The setup is similar to the one in Sec.~\ref{sec4}. The main difference is that the expanding particles now reach the boundaries of the lattice and equilibrate by moving back and forth in a finite system. The stationary quasimomentum distribution function in this case is not the rapidity distribution. Also, after particles reach the lattice boundaries, the emergent eigenstate solution is no longer valid~\cite{vidmar_iyer_17}.

\begin{figure}[!b]
\begin{center}
\includegraphics[width=0.99\columnwidth]{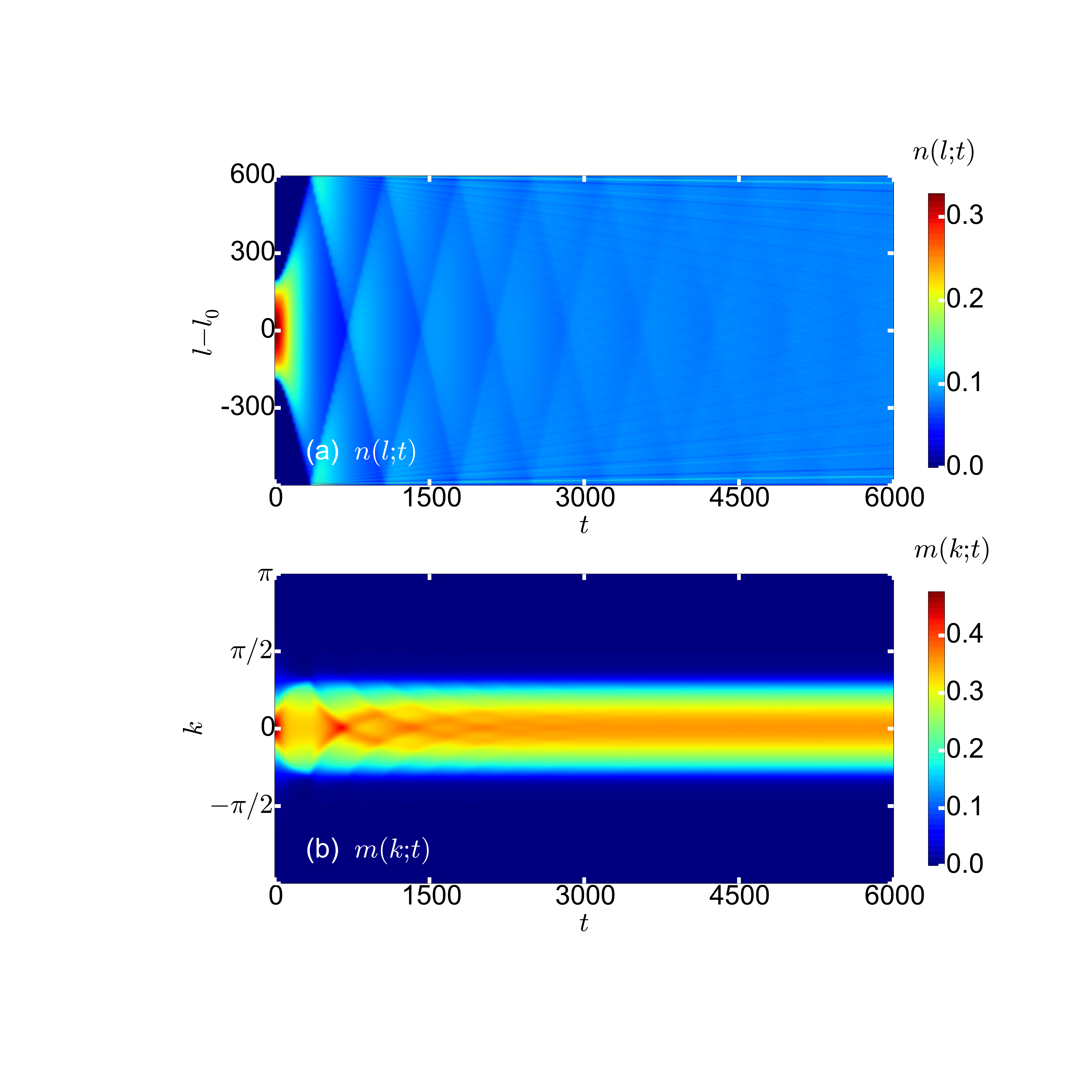}
\caption{Long-time dynamics of $N_p=100$ impenetrable SU($N$) fermions after turning off a harmonic trap with $R=200$ in a lattice with $L=1200$ sites ($l_0=600$). (a) Total site occupations $n(l;t)$. (b) Total quasimomentum distribution $m(k;t)$, computed following Eq.~(\ref{eq:nk}).} \label{fig9}
\end{center}
\end{figure}

Figure~\ref{fig9}(a) shows the dynamics of the total site occupations $n(l;t)$. At long times, as expected, $n(l;t)$ equilibrates and becomes (nearly) homogeneous, $n(l;t) \to N_p/L$. In the light of the findings in Sec.~\ref{sec4}, the dynamics of the total quasimomentum distribution function $m(k;t)$, shown in Fig.~\ref{fig9}(b), is more remarkable. After approaching the rapidity distribution during the expansion, when the propagating fronts reach the lattice boundaries for the first time, $m(k;t)$ begins to change and a revival occurs in the occupation of low quasimomentum modes. Ultimately, after some oscillations with decreasing amplitude, $m(k;t)$ equilibrates to a new distribution, which is different from that of the spinless fermions to which the impenetrable SU($N$) fermions are mapped.

\begin{figure}[!t]
\begin{center}
\includegraphics[width=0.99\columnwidth]{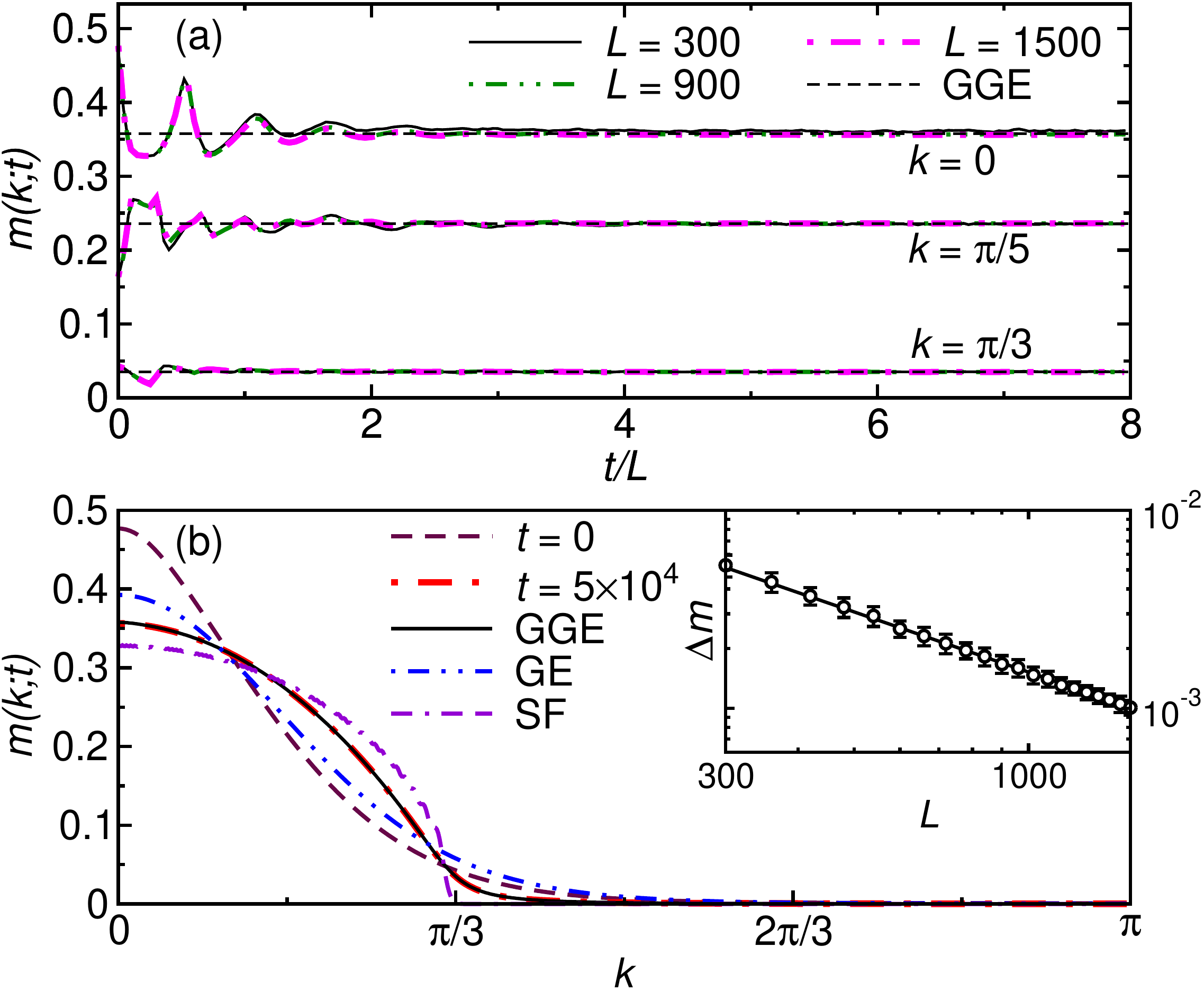}
\caption{Generalized thermalization in a box trap after turning off a harmonic trap. (a) Evolution of the total occupation of three quasimomentum modes ($k=0$, $\pi/5$, and $\pi/3$) plotted vs $t/L$ for three sizes $L$ of the box trap. For $L=1500$ there are $N_p=125$ impenetrable SU($N$) fermions and $R=250$. The other system sizes have the same ratios $N_p/L$ and $N_p/R$. The horizontal dashed lines are the GGE predictions for $L=1500$. (b) Total quasimomentum distribution function $m(k;t)$ in the initial state ($t=0$), in the time-evolving state after a long equilibration time ($t=50000$), in the GGE, in the grand canonical ensemble (GE), and for the spinless fermions (SF) onto which we map the impenetrable SU($N$) fermions. The results shown are for the $L=1500$ case in (a). (Inset) Distance $\Delta m$, see Eq.~(\ref{GGE_diff}), between time-averaged results from the time evolution and the GGE predictions. The time average is computed using 100 times between $t_\text{min}=10 L$ and $t_\text{max}= L^2/14$ (set to avoid including revivals that occur in timescales $\propto L^2$), with a time spacing $dt=(t_\text{max}-t_\text{min})/100$. Error bars show the standard deviation for the average. The solid line is a power law fit $\Delta m\propto L^{-\alpha}$ for $L\geq720$, which yields $\alpha=1.0(1)$.}\label{fig10}
\end{center}
\end{figure}

In Fig.~\ref{fig10}(a), we show the evolution of the total occupation of three quasimomentum modes ($k=0,\,\pi/5$, and $\pi/3$), as functions of time (normalized using the size of the final lattice $L$), for three values of $L$. We select $N_p$ and $R$ for the initial harmonic trap such that $N_p/L$ and $N_p/R$ are the same in all cases. Figure~\ref{fig10}(a) shows that, as expected, the time scale for the oscillations in $m(k;t)$ is set by $L$ (notice the data collapse for different values of $L$). Figure~\ref{fig10}(a) also makes apparent that, for $t/L\gtrsim 3$, the total occupations of the modes shown, and $m(k;t)$ in general, are nearly time independent, i.e., they have equilibrated.

In Fig.~\ref{fig10}(b), we show the initial total quasimomentum distribution function, the long-time equilibrated $m(k;t)$ (corresponding spin-resolved results are shown in Appendix~\ref{app1}), and the quasimomentum distribution of the spinless fermions to which we map the impenetrable SU($N$) fermions. They can all be seen to be different, with $m(k=0;t=0)$ being greater than the total occupation of the zero quasimomentum mode after equilibration, which in turn is greater than the occupation of the zero quasimomentum mode for spinless fermions.

In order to describe $m(k;t)$ after equilibration, we construct the generalized Gibbs ensemble (GGE)~\cite{rigol_dunjko_07, vidmar16}, whose density matrix is defined as
\begin{equation}
\hat{\rho}_{\rm GGE}=\frac{1}{Z_{\rm GGE}}e^{-\sum_k \lambda_k\hat I_k}\,,
\end{equation}
where $\{\hat I_k\}$ are the conserved quantities [the occupations of the single-particle eigenstates of the noninteracting spinless fermions to which we map the impenetrable SU($N$) fermions, see Eq.~(\ref{def_Hsf})], $Z_{\rm GGE}=\Tr[e^{-\sum_k \lambda_k\hat I_k}]$ is the partition function, and $\{\lambda_k\}$ are fixed by the initial values of the conserved quantities, $\Tr[\hat\rho_{\rm GGE}\hat I_k]=\langle\Psi^I_{\rm SF}|\hat I_k|\Psi^I_{\rm SF}\rangle\equiv\langle\hat I_k\rangle^I$. From these conditions, it follows that $\lambda_k=\ln[(1-\langle\hat I_k\rangle^I)/\langle\hat I_k\rangle^I]$, and that $Z_{\rm GGE}=\prod_k(1-\langle\hat I_k\rangle^I)^{-1}$ \cite{rigol_dunjko_07, vidmar16}.

Computing total one-body correlation functions of impenetrable SU($N$) fermions in the GGE is straightforward using the grand canonical ensemble approach introduced in Ref.~\cite{zhang18}. The off-diagonal part of the total one-body correlations in the GGE is given by
\begin{align}
C^{\rm GGE}_l(x\neq0)=\frac{1}{Z_{\rm GGE}}\{&\det[\mathbf{I}+(\mathbf{I}+\mathbf{A})\mathbf{M}_{l,x}\mathbf{U}^\dagger e^{-\mathbf{\Lambda}}\mathbf{U}]\nonumber\\&-\det[\mathbf{I}+\mathbf{M}_{l,x}\mathbf{U}^\dagger e^{-\mathbf{\Lambda}}\mathbf{U}]\}\,,
\end{align}
where $\bf I$ is the identity matrix, $\bf A$ is a matrix with elements $A_{ij} = \delta_{il}\delta_{j(l+x)}$, $\bf\Lambda$ is a diagonal matrix whose elements are the Lagrange multipliers $\lambda_k$, and $\mathbf{M}_{l,x}$ is the matrix form of the projection operator ${\cal \hat M}_{l,x}$ [see Eq.~(\ref{def_projM})], which is a diagonal matrix in which the elements between $l+1$ and $l+x-1$ are zero and the other ones are one. Moreover, $\bf U$ is the unitary matrix that diagonalizes $\hat H_{\rm SF}$ in Eq.~(\ref{def_Hsf}), where $\mathbf{U}\mathbf{H}_{\rm SF}\mathbf{U}^\dagger=\mathbf{E}$ (here $\mathbf{E}$ is a diagonal matrix with the single-particle eigenenergies). The diagonal part of the total one-body correlations in the GGE is given by
\begin{equation}
C^{\rm GGE}_l(x=0)=1-[\mathbf{I}+\mathbf{U}^\dagger e^{-\mathbf{\Lambda}}\mathbf{U}]^{-1}_{ll}\,,
\end{equation}
and is identical to that of the spinless fermions.

The results in Fig.~\ref{fig10}(a) show that the total occupation of the quasimomentum modes of the impenetrable SU($N$) fermions equilibrate at the values predicted by the GGE. In Fig.~\ref{fig10}(b), the long-time result for $m(k;t)$ is indistinguishable from the GGE one, and they are both clearly different from the prediction of the grand canonical ensemble (GE)~\cite{zhang18} in which the temperature and the chemical potential are set by the energy and the number of particles in the system~\cite{dalessio_kafri_16}. This makes apparent that these systems do not equilibrate at the traditional statistical mechanics prediction (do not thermalize).

In order to understand how the differences between the GGE and the exact dynamics predictions behave as one increases the system size, we compute the following time average of the differences between $m(k;t)$ after equilibration and the GGE prediction:
\begin{equation}
\label{GGE_diff}
\Delta m=\frac{1}{M}\sum_{i=1}^M\frac{\sum_{k}|m(k,t_i)-m_{\rm GGE}(k)|}{\sum_{k}m_{\rm GGE}(k)}.
\end{equation}
Results for $\Delta m$ vs $L$ are reported in the inset in Fig.~\ref{fig10}(b). They show that 
$\Delta m\propto L^{-1}$, consistent with findings in studies of 1D hard-core bosons~\cite{gramsch_rigol_12, wright_rigol_14}. These results suggest that observables after long times and the GGE predictions become identical in the thermodynamic limit ($L\to\infty$, with $N_p/L$ and $N_p/R$ fixed).

\begin{figure}[!bt]
\begin{center}
\includegraphics[width=0.99\columnwidth]{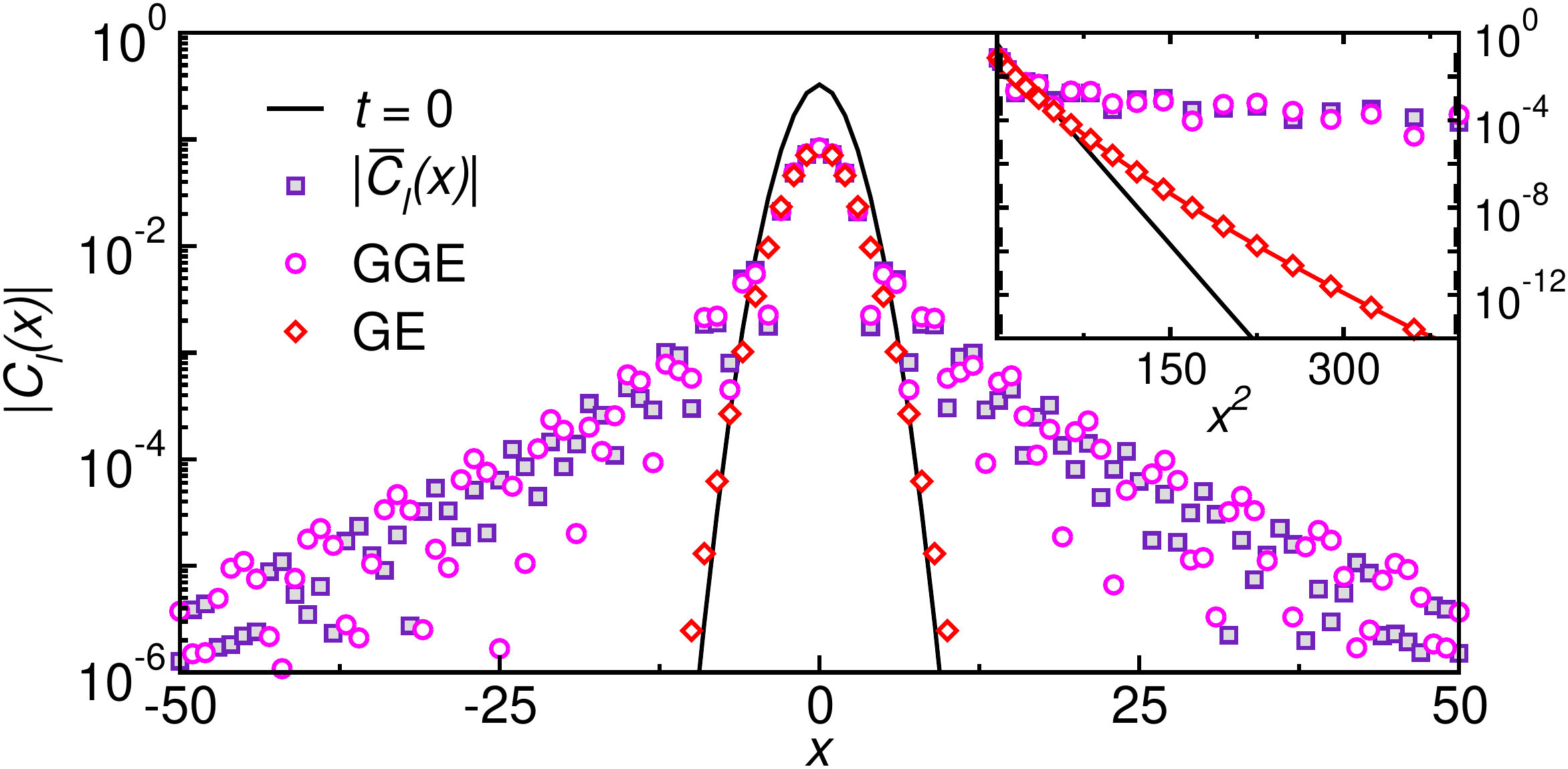}
\caption{Total one-body correlations $|C_l(x)|$ after equilibration. Comparison between the time-averaged total one-body correlations $|\overline{C}_l(x)|$ after equilibration [see Eq.~(\ref{GGE_diff_Cl})], the initial total one-body correlations, and the predictions of the GGE and the grand canonical ensemble (GE). $\overline{C}_l(x)$ is computed using times $t_i$ in the interval $15000 \leq t_i\leq 160000$, with a time spacing $dt=50$. The system studied has $L=1500$, $N_p=125$, and initial $R=250$. (Inset) Same results plotted versus $x^2$.}\label{fig11}
\end{center}
\end{figure}

Finally, we discuss the long-time behavior of the total one-body correlations for the largest system size considered in Fig.~\ref{fig10} ($L=1500$). For every $x$, we define the time average
\begin{equation} \label{GGE_diff_Cl}
\overline{C}_l(x) = \frac{1}{M} \sum_{i=1}^MC_l(x;t_i).
\end{equation}
Results for $\overline{C}_l(x)$ are shown in Fig.~\ref{fig11} for $l=L/2$. They reveal that during the equilibration process the Gaussian decaying correlations present in the initial state evolve toward a slower, exponential-like, decay at large $x$. Those correlations are well described by the GGE, and are very different from the stretched exponential decay predicted by the grand canonical ensemble~\cite{zhang18}.

\section{Summary} \label{sec_conclusion}

We introduced an exact approach to study the time evolution of spin-resolved one-body observables after quantum quenches in the (generalized) SU($N$) Fermi-Hubbard model at infinite repulsion. This approach is tailored for initial states that exhibit generalized Neel order, namely, in which there is a periodic $N$-spin pattern with consecutive fermions carrying distinct spin flavors. Our approach is based on a compact representation of charge degrees of freedom as spinless fermions and spin degrees of freedom as nonlocal string of operators (projectors) expressed in terms of spinless fermions. It can be used to benchmark numerical and analytical studies of quantum quenches involving (generalized) Neel order in the (generalized) Fermi-Hubbard model in the limit of very strong repulsive interactions.

We studied three unique phenomena that occur during expansion dynamics of impenetrable SU($N$) fermions far from equilibrium. They are within reach in current optical-lattice experimental setups. The first one occurs during the melting of generalized Neel domain walls, in which we unveiled a dynamical emergence of Gaussian (ground-state-like~\cite{zhang18}) correlations in the resulting current-carrying steady state. We explained this phenomenon using an emergent eigenstate solution to quantum dynamics~\cite{vidmar_iyer_17}. The second one occurs during the sudden expansion of harmonically trapped impenetrable SU($N$) fermions into an empty lattice after turning off the trap. We showed that the total quasimomentum distribution function of the impenetrable SU($N$) fermions evolves toward the quasimomentum distribution function of the spinless fermions onto which we map the former. Namely, as seen in other integrable systems~\cite{rigol05, vidmar_xu_17, minguzzi05, buljan_pezer_08, jukic_klajn_09, iyer12, campbell15, piroli_calabrese_17, sutherland_98, bolech12, campbell15, mei16}, the quasimomentum distribution function of the real particles transforms into the rapidity distribution at long expansion times. Finally, we studied the equilibration dynamics of impenetrable SU($N$) fermions in a box trap after turning off a harmonic trap. We showed that observables (such as the quasimomentum distribution function) after equilibration can be described using a GGE.  

\section{Acknowledgments}

We acknowledge discussions with B.~Bertini and W.~de Roeck. Y.Z.~and M.R.~acknowledge support from NSF Grant No.~PHY-1707482. L.V.~acknowledges support from the Slovenian Research Agency (ARRS), research core funding No.~P1-0044.


\appendix


\section{Spin-resolved quasimomentum distribution function} \label{app1}

In the main text, in order to minimize finite-size effects, we focused on total one-body correlations [defined in Eq.~\eqref{def_C_integrated}] and the corresponding total quasimomentum distribution functions. A detailed study of finite-size effects in spin-resolved one-body correlations and quasimomentum distribution functions of impenetrable SU($N$) fermions in equilibrium was presented in Ref.~\cite{zhang18}. There we showed that those spin resolved observables approach the averaged ones (the total ones divided by the number of spin flavors) as one increases the number of particles while keeping the number of spin flavors fixed.

\begin{figure}[!t]
\begin{center}
\includegraphics[width=0.99\columnwidth]{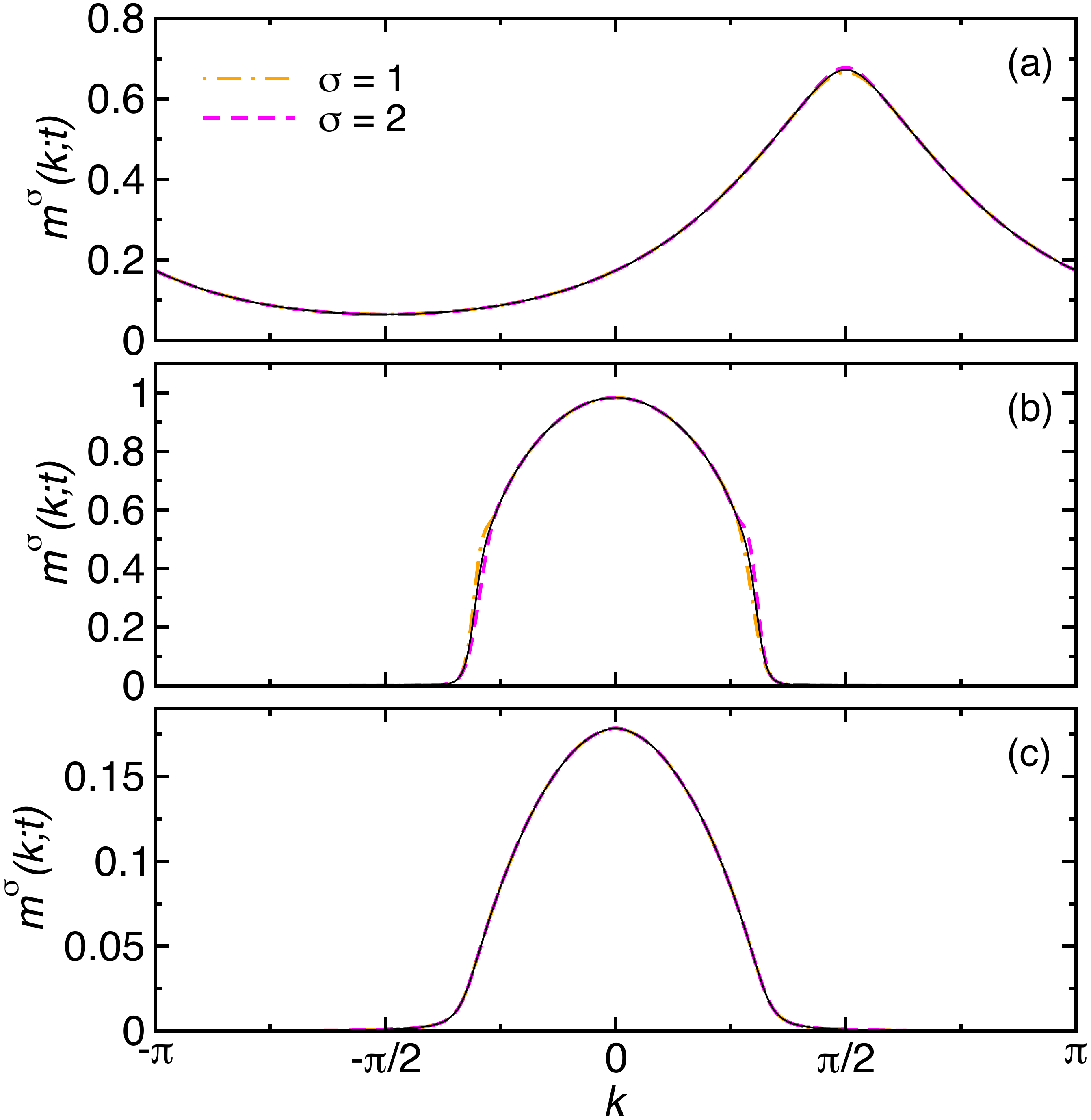}
\caption{Spin-resolved quasimomentum distribution function $m^{\sigma}(k;t)$ for impenetrable SU(2) fermions with generalized Neel order. (a) Melting of a domain wall with $N_p=400$ fermions at $t=200$ [see Fig.~\ref{fig5}(a) in Sec.~\ref{sec3}]. (b) Sudden expansion of $N_p=100$ fermions from a harmonic trap with $R=200$ at $t=375$ [see Fig.~\ref{fig7}(a) in Sec.~\ref{sec4}]. (c) Equilibration of $N_p=100$ particles in a box trap with $L=1200$. Results are shown for $t=50\,000$ after turning off a harmonic trap with $R=200$ [see Fig.~\ref{fig10}(b) in Sec.~\ref{sec5}]. Thin black lines depict the results averaged over spin flavors $\bar{m}(k;t)=\sum^{2}_{\sigma=1}m^{\sigma}(k;t)/2$. }\label{figapp1}
\end{center}
\end{figure}

A similar study of finite-size effects in spin-resolved one-body correlations and quasimomentum distribution functions out of equilibrium is beyond the scope of this work. As for the equilibrium systems studied in Ref.~\cite{zhang18}, we find that those observables out of equilibrium approach the averaged ones when increasing the number of particles (for a fixed number of spin flavors). Finite-size effects are small for the system sizes considered in the main text. In Fig.~\ref{figapp1}, we show the spin-resolved quasimomentum distribution functions $m^{\sigma}(k;t)$ for impenetrable SU(2) fermions ($\sigma=1,2$). Results for the melting of the generalized Neel domain wall (Sec.~\ref{sec3}), the sudden expansion after turning off the harmonic trap (Sec.~\ref{sec4}), and equilibration in the box trap (Sec.~\ref{sec5}), are shown in Figs.~\ref{figapp1}(a),~\ref{figapp1}(b), and~\ref{figapp1}(c), respectively. The results for individual spin flavors are almost indistinguishable from the averaged ones, which are the total ones divided by 2.

\bibliographystyle{biblev1}
\bibliography{references}

\end{document}